\documentclass[aps,pre,twocolumn,a4paper,10pt]{revtex4-1}%
\usepackage[hidelinks,colorlinks=false]{hyperref}
\usepackage{amssymb}
\usepackage{amsmath}
\usepackage{mathrsfs}
\usepackage{bm}
\usepackage[dvipsnames]{xcolor}
\usepackage[sort&compress]{natbib}
\usepackage{graphicx}
\usepackage{color}

\usepackage[utf8]{inputenc}

\renewcommand{\vec}[1]{\bm{#1}}

\newcommand{\restrdelta}{\left(\left.\Delta\right|_\gamma\right)}

\newcommand{\norm}[1]{\left|#1\right|}

\newcommand{\dmi}[1]{\textcolor{black}{#1}}%

\begin{document}

\title{Magnetic Polymer Models for Epigenetics-Driven Chromosome Folding}

\author{Davide Col\`i}
\thanks{Joint first author}
\affiliation{Dipartimento di Fisica and Sezione INFN, Università degli Studi di Padova, I-35131 Padova, Italy}
\author{Davide Michieletto$^*$}
\thanks{Joint first author}
\email{davide.michieletto@ed.ac.uk}
\affiliation{SUPA, School of Physics and Astronomy, University of Edinburgh, Edinburgh EH9 3FD, UK}
\affiliation{MRC Human Genetics Unit, Institute of Genetics and Molecular Medicine, University of Edinburgh, Edinburgh EH4 2XU, UK}
\affiliation{Centre for Mathematical Biology, and Department of Mathematical Sciences, University of Bath, North Rd, Bath BA2 7AY, UK}
\author{Davide Marenduzzo} 
\affiliation{SUPA, School of Physics and Astronomy, University of Edinburgh, Edinburgh EH9 3FD, UK}
\author{Enzo Orlandini}
\email{orlandini@pd.infn.it}
\affiliation{Dipartimento di Fisica and Sezione INFN, Università degli Studi di Padova, I-35131 Padova, Italy}
%\date{\today}

\begin{abstract}
\vspace{-0.3 cm}
\textbf{Epigenetics is a driving force of important and ubiquitous phenomena in nature such as cell differentiation or even metamorphosis. Oppositely to its widespread role, understanding the biophysical principles that allow epigenetics to control and rewire gene regulatory networks remains an open challenge. In this work we study the effects of epigenetic modifications on the spatial folding of chromosomes -- and hence on the expression of the underlying genes -- by mapping the problem to a class of models known as magnetic polymers. 
In this work we show that a first order phase transition underlies the simultaneous spreading of certain epigenetic marks and the conformational collapse of a chromosome. Further, we describe Brownian Dynamics simulations of the model in which the topology of the polymer and thermal fluctuations are fully taken into account and that confirm our mean field predictions. 
Extending our models to allow for non-equilibrium terms yields new stable phases which qualitatively agrees with observations \emph{in vivo}.
Our results show that statistical mechanics techniques applied to models of magnetic polymers can be successfully exploited to rationalize the outcomes of experiments designed to probe the interplay between a dynamic epigenetic landscape and chromatin organization.
}
\vspace{-1 cm}
\end{abstract}

\maketitle

\section{Introduction}
Epigenetic effects in biology are defined as inheritable changes to a phenotype that do not involve alterations in the underlying DNA sequence~\cite{Waddington1942a,Alberts2014,Cortini2015}. The best and perhaps most important example of this class of changes is that resulting in the difference between cell types in our body: while all cells possess the same DNA, they can differentiate into neurons, epithelium, retinal, etc. Other important examples in which epigenetic effects are at work are the reprogramming of germ~\cite{Tang2015a} and pluripotent~\cite{Rulands2018} cells, temperature-dependent sex determination in certain fish and reptiles~\cite{Crews2003}, metamorphosis of certain animals (such as caterpillars into butterflies) and generic polyphenism in insects~\cite{Ernst2015a}. All these widespread phenomena share the fact that while the original gene regulatory networks are re-wired to give rise to previously absent features, organs or appendices, the underlying DNA sequence is untouched.
 
It is becoming increasingly clear that epigenetic changes are orchestrated through certain biochemical -- or epigenetic -- marks that are deposited along the genome. How these marks affect gene expression is, however, still poorly understood. One widespread idea is that epigenetic marks change the spatial folding of the genome and hence its accessibility for proteins and transcription factors; in turn, this change in local folding affects the expression of the underlying genes~\cite{Grewal2011}. Increasing amount of evidence shows that genome folding correlates with epigenetics~\cite{Boettiger2016} but the biophysical mechanisms controlling this folding are unclear, especially considering that epigenetic marks are transient biochemical modifications which are dynamically deposited to and removed from the substrate~\cite{Rulands2018,Cortini2015}. 

\begin{figure}[!t]
	\centering
	\includegraphics[width=0.45\textwidth]{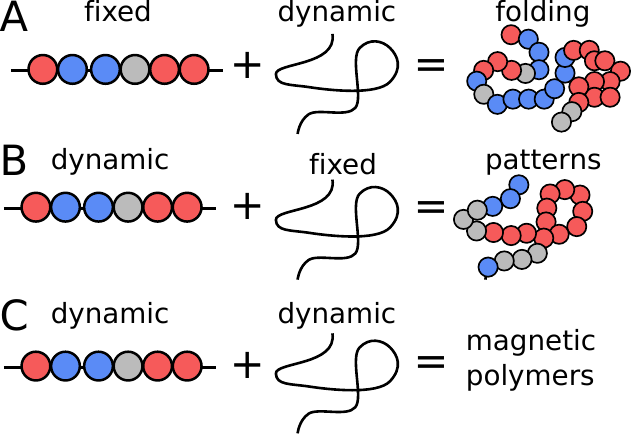}
	\vspace{-0.3 cm}
	\caption{\textbf{Classes of Models for Epigenetic Patterns and Chromosome Folding.} \textbf{A} Co-polymer~\cite{Jost2014B,Barbieri2012} and bridging-induced attraction~\cite{Brackley2013pnas,Brackley2016nar,Brackley2016genomebiol} models predict the folding of a chromosome given a certain fixed pattern of epigenetic marks.  \textbf{B} Ising-like models predict the spreading and patterns of epigenetic marks given a passive substrate with a given average contact probability~\cite{Dodd2007}. \textbf{C} In this work we study magnetic polymer models that couple the dynamics of epigenetic marks with that of the polymeric substrate. }
	\vspace{-0.5 cm}
	\label{fig:model}
\end{figure}

First, several models have been proposed to explain how a chromosome may fold given a certain \emph{fixed} pattern of epigenetic marks: among the most successful are the bridging-induced attraction~\cite{Brackley2013pnas,Brackley2016nar,Brackley2016genomebiol}, the strings-and-binders~\cite{Barbieri2012} and co-polymer~\cite{Jost2014B} models (Fig.~\ref{fig:model}A). Second, a different class of models have been developed to understand how epigenetic marks may spread and self-organise into patterns on a \emph{passive} substrate with a given average probability of looping~\cite{Dodd2007,Micheelsen2010,Jost2014pre,Berry2017} (Fig.~\ref{fig:model}B). Both classes of models lack an element that is natural to consider in a more complete and refined model: the interplay between the kinetics of the epigenetic marks and the dynamical spatial arrangements of the polymeric substrate (see Fig.~\ref{fig:model}C). Accounting for this interplay maps the problem to that of ``magnetic polymer'' or ``annealed co-polymer'' models~\cite{Garel1999a}.
\dmi{So far few groups have developed polymer models apt to specifically address questions concerning  chromatin folding~\cite{Michieletto2016prx,Michieletto2017scirep,Michieletto2018nar,Haddad2017, Jost2018} but there are still many open questions that can be addressed within a more physically-minded statistical mechanics framework.}

%%%%%%%%%%%%%%%% begin figure %%%%%%%%%%%%%%%%%%%
%%%%%%%%%%%%%%%% end figure %%%%%%%%%%%%%%%%%%% 

In this, and the companion~\cite{PRL}, work we focus on field-theoretic approaches to study analytically and numerically solvable models of magnetic polymers. \dmi{This paper is organised as follows. In Section~\ref{sec:eqmodel} we derive a free energy for a single chromosome with epigenetic and 3D dynamics from the partition function of a magnetic polymer. In Section~\ref{sec:noneqmodel} we derive the kinetics equations for the equilibrium model and generalise them to described non-equilibrium conditions. In Sections~\ref{sec:Brownian_model} and ~\ref{sec:Brownian_single_chrom} we detail the computational strategy that we use to perform Brownian dynamics simulations of coarse-grained polymers with dynamic (epigenetic) ``recolouring'' and we explain how to break detailed balance and achieve non-equilibrium conditions (also explored in Ref.~\cite{Michieletto2016prx}). Finally, in Section~\ref{sec:Brownian_melts} we discuss the 
Brownian dynamics simulations of melts of polymers with recolouring dynamics.}
By studying these models we show that equilibrium models are not compatible with experiments and a qualitative agreement is recovered by accounting for non-equilibrium processes. We show that non-equilibrium effects can either promote new phases or stabilise an arrested phase separated coexistence of epigenetic marks. 

\section{Equilibrium Model } 
\label{sec:eqmodel}
We describe a single chromosome as an $N$-step self-avoiding walk (SAW) on a lattice with coordination number $z$ and with each vertex $i$ displaying an epigenetic state, or spin, $q_i$.  \dmi{A generic situation is one in which two classes of epigenetic states (plus one neutral) are present: those marking transcriptionally active (say $q=1$) and inactive (say $q=-1$) chromatin~\cite{Cortini2015,Lieberman-Aiden2009}. A neutral state ($q=0$) may be associated to a region of chromatin that does not contain any clear epigenetic mark, such as gene deserts in Drosophila~\cite{Filion2010}.} Non-neutral states are known to associate with specific protein complexes that can mediate chromosome bridging through ``reading'' domains (e.g. HP1~\cite{Larson2017}, PRC~\cite{Pinter2012} or others~\cite{Barbieri2012,Brackley2013pnas,Jost2014B}) and epigenetic spreading through ``writing'' domains (e.g. Suv39~\cite{Hathaway2012,Michieletto2016prx}). 

Within this magnetic polymer framework, both processes (bridging and spreading) can be modelled by the same ferromagnetic interaction that tends to align 3D proximal spins, or tends to bring them together when already aligned~\cite{Michieletto2016prx,Michieletto2018nar,Michieletto2017scirep}.
Thus, any pair of 3D proximal monomers interact with each other via a contact potential that depends on the values of $q$ in a pair.  More precisely, if the $i$-th and the $j$-th monomers are nearest neighbours on the lattice, their contact energy $J(q_i, q_j )$ is
\begin{small}
\begin{equation}\label{eq:contactenergy}
 J(q_i, q_j) = 
	      \begin{cases}
		      -\epsilon& \text{ if } q_i = q_j = \pm 1\\
		      0 & \text{ otherwise} 
	       \end{cases} \quad,
\end{equation}
\end{small}
with $\epsilon > 0$. Note that the mark $q=0$ does not contribute to this configurational energy and we will define it as a neutral mark. The equilibrium properties  of this system are described by the partition function 
\begin{equation}\label{eq:partitionmarks}
 \mathcal{Z} = \sum_{\text{SAW}} \sum_{\{q\}} \exp\left[-\frac{\beta}{2} \sum_{i,j=1}^N \Delta_{\vec{r}_i ,\vec{r}_j} J(q_i, q_j)
 \right],
\end{equation}
where $1/\beta= k_B T$. The sums $\sum_{\text{SAW}}$ and $\sum_{\{q\}}$ run over the set of all $N$-steps SAWs and all the possible combinations of $N$ epigenetic states, respectively. 
The matrix $\Delta_{\vec{r}_i \vec{r}_j}$ is the 
adjacency matrix associated to a given SAW and is given by
\begin{equation}
 \Delta_{\vec{r}_i \vec{r}_j} = 
	\begin{cases}
	      1 & \text{if $\vec{r}_i$ and $\vec{r}_j$ are nearest neighbours in 3D} \\
	      0 & \text{otherwise}
	\end{cases} . 
\end{equation}

Notice that the partition function in Eq.~\eqref{eq:partitionmarks} presents a clear $\mathbb{Z}_2$ symmetry as $J(-q_i, -q_j) = J(q_i, q_j)$.

%\dmi{Since we are here interested in the critical properties of the system, we can ict the phase space of the epigenetic variables, $q$, to the case where the abundance of the state $q=0$ is equal to the one of $q=-1$. With this restriction the system can be faithfully described by a two-valued spin variable $S=\{1,-\tfrac12\}$ where $S=1$ corresponds to the mark $q=1$, while the values $S=-\tfrac12$ has multiplicity 2 as it corresponds both to $q=0$ and $q=-1$~\cite{Wu1982}.}
%\dmi{ XXX ENZO XXX}

Since here we focus on the transitions between a disordered and an ordered phase, irrespective of the mark type, we follow the strategy, sometimes  used in Potts models~\cite{Wu1982}, to restrict the phase space of the epigenetic variables, $q$, to the case where the states $q=0$ and $q=-1$ are equally populated. With this restriction the $\mathbb{Z}_2$ symmetry of Eq.~\eqref{eq:partitionmarks} is violated but the system can be faithfully described by a two-valued spin variable $S=\{1,-\tfrac12\}$  where $S=1$ corresponds to the mark $q=1$, while the values $S=-\tfrac12$ has multiplicity $2$ as it corresponds both to $q=0$ and $q=-1$. .

By using the spin variable $S$, the coupling in Eq.~\eqref{eq:contactenergy} becomes
\begin{small}
\begin{equation}
 J(S_i, S_j) =
	      \begin{cases}
		      -\epsilon& \text{ if } S_i = S_j = 1\\
		      -\tfrac14\epsilon& \text{ if } S_i = S_j = -\tfrac12 \\
		      0 & \text{ otherwise} 
	       \end{cases} \quad.
\end{equation}
\end{small}
which can be re-written as
\begin{small}
\begin{equation}\label{eq:contactenergy2}
  J(S_i, S_j) = -\tfrac59\epsilon \left(S_i+\tfrac15\right)\left(S_j+\tfrac15\right) - \tfrac15\epsilon \, .
\end{equation}
\end{small}
The partition function in Eq.~\eqref{eq:partitionmarks} can then be recast into the following form,
\begin{small}
\begin{equation} \label{eq:partitionspin}
\mathcal{Z} = \sum_{\text{SAW}} \sum_{\{S\}} \exp\left[\tfrac{\epsilon \beta }{2}\sum_{i,j=1}^N  
\Delta_{\vec{r}_i\vec{r}_j} \left(\tfrac59(S_i +\tfrac15)(S_j +\tfrac15)+ \tfrac15\right)\right]\,\,.
\end{equation}
\end{small}
We now perform a  Hubbard-Stratonovich transformation to rewrite the term
\begin{small}
\begin{equation}\label{eq:firstterm}
\exp\left[\frac{5\epsilon \beta }{18}\sum_{i,j=1}^N (S_i +\tfrac15)\Delta_{\vec{r}_i\vec{r}_j} (S_j +\tfrac15)\right] 
\end{equation}
\end{small}
as 
\begin{small}
%\begin{equation*}
%\int {\cal{D}} \vec{\phi} \exp\left[- \tfrac{9}{10 \epsilon \beta} \sum_{i,j=1}^N \phi_i \Delta_{\vec{r}_i\vec{r}_j}^{-1} \phi_j 
%+\sum_{i=1}^N \log\left(\sum_{\{S_i\}} e^{\phi_i(S_i+\tfrac15)}\right)\right]
%\end{equation*}
\begin{equation*}
\int {\cal{D}} \vec{\phi} \exp\left[- \tfrac{9}{10 \epsilon \beta} \sum_{i,j=1}^N \phi_i \Delta_{\vec{r}_i\vec{r}_j}^{-1} \phi_j 
+\sum_{i=1}^N \phi_i(S_i+\tfrac15) \right]
\end{equation*}
\end{small}
\dmi{where ${\cal{D}}\vec{\phi} = (2\pi)^{-N/2} (9/5\epsilon\beta)^{1/2} (\det \Delta^{-1} )^{1/2} \prod_{i=1}^N \mathrm{d} \phi_i$}. By summing over all possible spin configurations $\{ S \}$ we get
\begin{small}
\begin{align}
\int {\cal{D}}\vec{\phi} \exp\left[- \tfrac{9}{10 \epsilon \beta} \sum_{i,j=1}^N \phi_i \Delta_{\vec{r}_i\vec{r}_j}^{-1} \phi_j \right] \prod_{i=1}^N \left( e^{ \tfrac65 \phi_i} + 2e^{-\tfrac{3}{10}\phi_i} \right) \notag 
\end{align}
\end{small}
which can be re-written as 
\begin{small}
\begin{align}
\int {\cal{D}}\vec{\phi} \exp\left[- \tfrac{9}{10 \epsilon \beta} \sum_{i,j=1}^N \phi_i \Delta_{\vec{r}_i\vec{r}_j}^{-1} \phi_j  + \sum_{i=1}^N \log{\left( e^{ \tfrac65 \phi_i} + 2e^{-\tfrac{3}{10}\phi_i} \right)} \right] \, .
\end{align}
\end{small}
This integral can be evaluated through a homogeneous saddle point approximation, \dmi{i.e. by replacing the integral with the value of the integrand at its maximum which we assume is attained for $\phi_i=\phi$. This approximation yields}
\begin{small}
\begin{equation}\label{eq:hubb1}
\mathcal{Z} = \sum_{\text{SAW}}
A \exp\left[- \tfrac{9}{10 \epsilon \beta} \phi^2 \sum_{i,j=1}^N  \Delta_{\vec{r}_i\vec{r}_j}^{-1} + N  \log\left(e^{\tfrac65\phi}+2e^{-\tfrac{3}{10}\phi}\right)\right] \,
\end{equation}
\end{small}
\dmi{where $A= (2\pi)^{-N/2} (9/5\epsilon\beta)^{1/2} (\det \Delta^{-1} )^{1/2}$}.  

In general, the term $ \sum_{i,j=1}^N  \Delta_{\vec{r}_i\vec{r}_j}^{-1}$, depends on the given SAW and it is not easy to compute. However, it can be estimated if we restrict the set of SAWs to the ones that are almost space filling, such as for example the subset of  Hamiltonian walks~\cite{Garel1999a}. 
A Hamiltonian walk is a path that visits each vertex of a lattice embedded in a volume $V$ exactly once and have been used to study equilibrium properties of highly compact polymers~\cite{Duplantier1987,Orland1985}. For a Hamiltonian walk, the adjacency matrix of the SAW $\Delta$ takes the same form of the adjacency matrix of the underlying lattice and it is characterised by the coordination number $z$. Hence, $\sum_{i,j=1}^N  \Delta_{\vec{r}_i\vec{r}_j}^{-1} = N/z$. Here, we consider $N$-steps configurations that, similarly to Hamiltonian walks, are contained in a volume $V$ but may in principle display a lower mean number of nearest neighbours, i.e. $\rho z$ instead of $z$. One can thus approximate 
\begin{equation}
 \sum_{i,j=1}^N  \Delta_{\vec{r}_i\vec{r}_j}^{-1} \approx \frac{N}{\rho z} \quad .
 \label{eq10}
\end{equation}
Note that for generic SAWs with low $\rho$, the right hand side in Eq.~\eqref{eq10} is actually an upper bound.
Finally, by following the approach described in Ref.~\cite{Nemirovsky1992} we  evaluate the last term in Eq.~\eqref{eq:partitionspin}, i.e.
\begin{small}
\begin{equation}
e^{F_{\text{SAW}}} = \sum_{\gamma \in \text{SAW}} 
\exp\left[\frac{\beta \epsilon}{10} \sum_{i,j=1}^N \restrdelta_{ij} \right]
\end{equation}
\end{small}
by approximating $F_\text{SAW}$ as
\begin{equation}\label{eq:FSAW}
\frac{F_\text{SAW}}{T N}  \approx - \log\left(\frac{z}{e}\right)+ \frac{1-\rho}{\rho}\log(1-\rho) - \frac{\beta \epsilon z}{10 }\rho\quad.
\end{equation}

\begin{figure}[t!]
 \centering
 \includegraphics[width=0.45\textwidth]{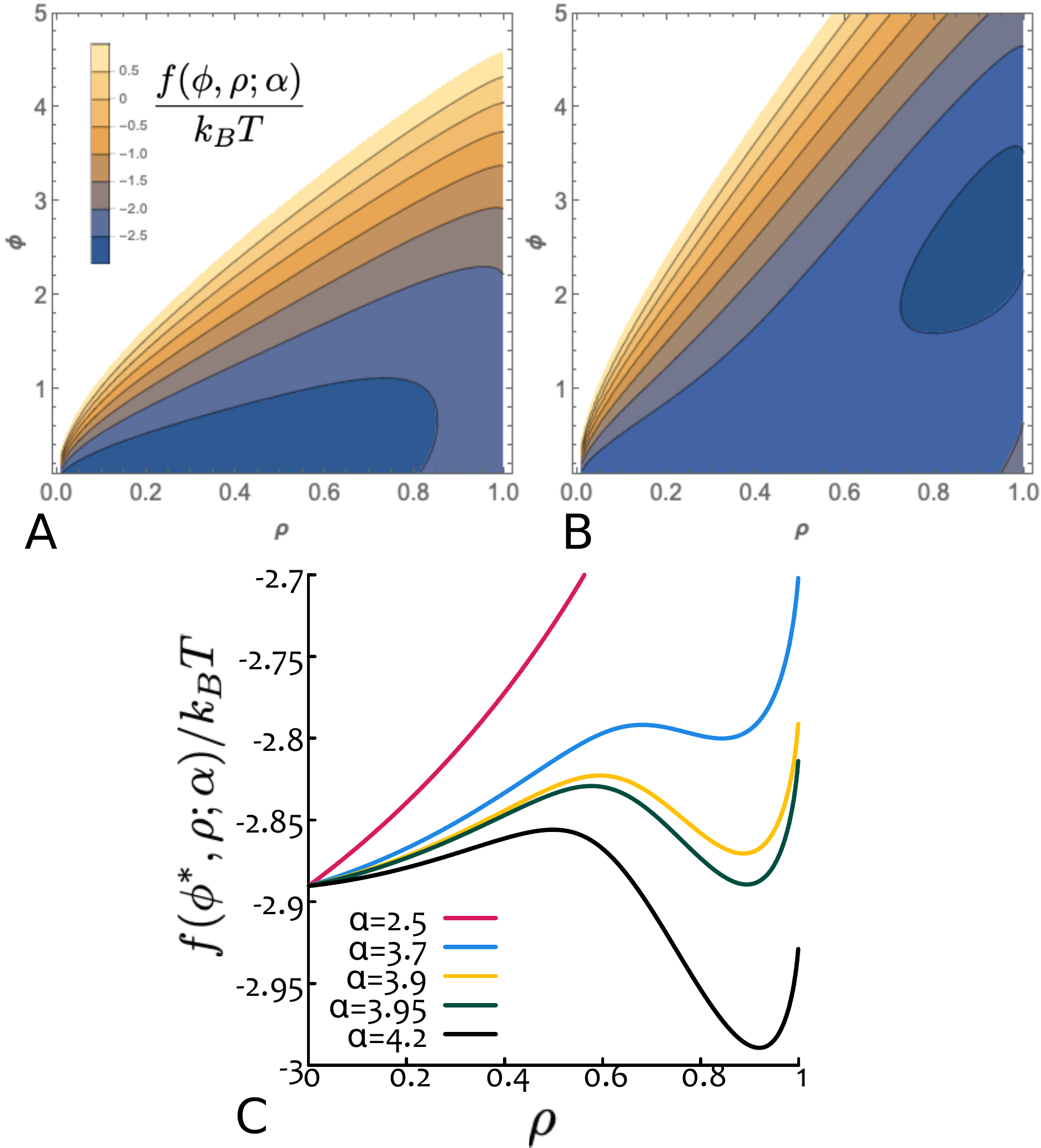}
 \caption{\textbf{Free Energy of Single Chromosome in Equilibrium.}  \textbf{A-B} Contour plots of the free energy in Eq.~\eqref{eq:freeenergysingle} for $\alpha=2.5$ and $\alpha=4$ respectively. \textbf{C} Plot of the minimised free energy $f(\phi^*,\rho;\alpha)$ as a function of $\rho$ and for different values of $\alpha$. One can notice that, at the critical value $\alpha_c \simeq 3.95$, there is a first order transition. } 
 \vspace{-0.5 cm}
 \label{fig:f_Single}
\end{figure}

By collecting all the terms and taking $f = -\frac{T}{N}\log\mathcal{Z}$, we obtain the following mean-field free energy density
\begin{equation}\label{eq:freeenergysingle}
\begin{split}
\frac{f}{T} =&   - \log\left(\frac{z}{e}\right)+ \frac{1-\rho}{\rho}\log(1-\rho) - \frac{\alpha}{10 }\rho + \\
& +  \frac{9}{10\alpha} \frac{\phi^2}{\rho}  - \log\left(e^{\tfrac65\phi}+2e^{-\tfrac{3}{10}\phi}\right) \dmi{- \frac1N \log{A}}
\end{split}\quad,
\end{equation}
where $\alpha \equiv \beta \epsilon z$ \dmi{and the term $\log{A}/N$ is negligible for $N \to \infty$}.
The equilibrium properties of the model are then obtained by finding the value of magnetisation and density that minimize Eq.~\eqref{eq:freeenergysingle}. First, the values of $\phi^*$ such that $\left. \partial f/\partial \phi \right|_{\phi^*} = 0$ are obtained by solving
\begin{equation}
\frac{9\phi}{5 \alpha \rho}+\frac{-\frac{3}{5} e^{-\frac{3\phi}{10}}+\frac{6}{5} e^{\frac{6\phi}{5}}}{2 e^{-\frac{3\phi}{10}}+e^{\frac{6\phi}{5}}} =0 \, .
\end{equation}
This equation can be solved numerically to find $\phi^*(\rho;\alpha)$ which can then be put back into the free energy to express it as $f(\phi^*,\rho;\alpha) = f^*(\rho;\alpha)$. This function is plotted in Fig.~\ref{fig:f_Single} for different values of $\alpha$ showing the appearance of a local minimum at $\rho >0$, which becomes a global minimum for $\alpha > \alpha_c$. 
The behaviour shown in Fig.~\ref{fig:f_Single} is characteristic of a first order transition: i.e., the value of $\rho^*$ minimising $f(\phi^*, \rho;\alpha)$ is discontinuous in $\alpha$ and jumps from $\rho^*=0$ to $\rho^*>0$ at $\alpha=\alpha_c \simeq 3.95$. This transition in density also corresponds to a discontinuous jump in magnetisation $\phi^*(\rho^*;a)$ from $\phi^*=0$ to $\phi^* \neq 0$. 

\begin{figure}[t!]
	\centering
	\includegraphics[width=0.35\textwidth]{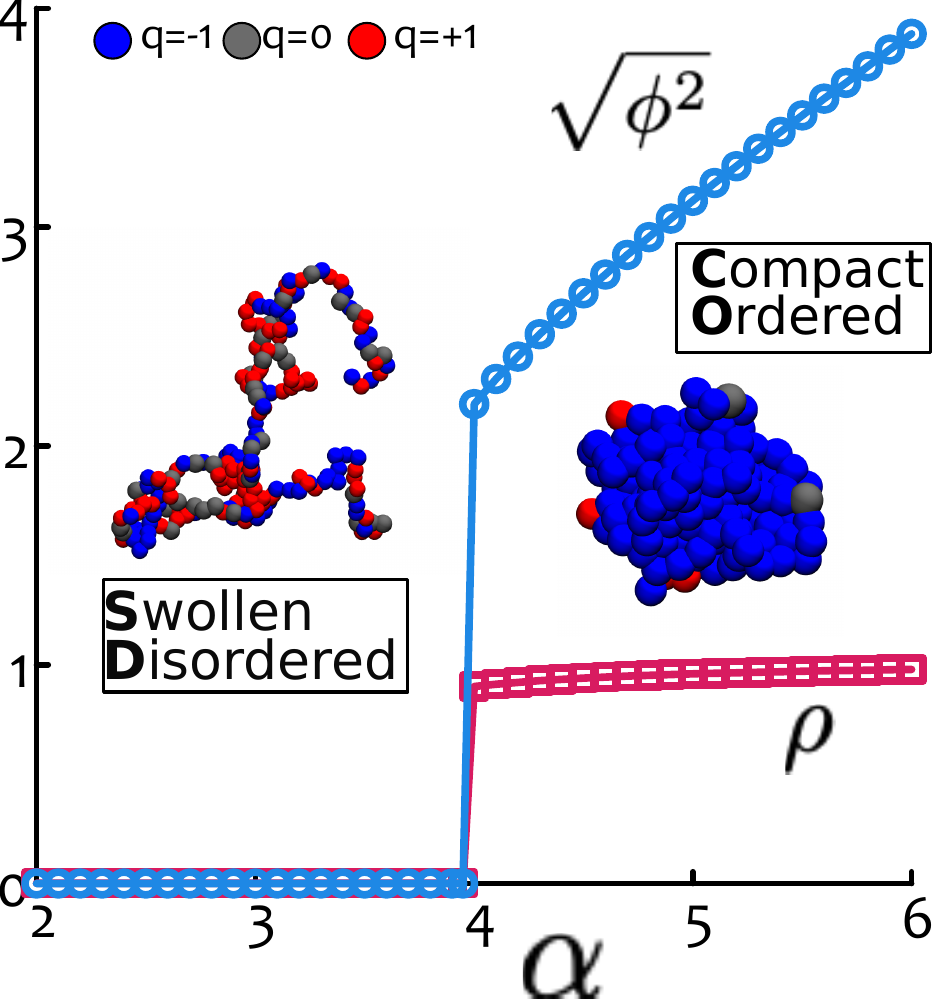}
	\vspace{-0.3 cm}
	\caption{\textbf{Equilibrium Phase Diagram for Single Chromosomes.} The phase diagram is obtained by numerical minimisation of Eq.~\eqref{eq:freeenergysingle} (red and blue symbols refer respectively to the order parameter and density profiles). The first order transition occurs at $\alpha = \epsilon/k_B T \simeq 3.95$ and separates a swollen-disordered (SD) phase from a compact-ordered (CO) one. The insets report snapshots from BD simulations at $\alpha_{BD}= 0.6$ and $\alpha_{BD}=1.1$ (the critical value of $\alpha$ for the model used in the BD simulations is $\alpha_{BD,c}\simeq 0.9$~\cite{Michieletto2016prx}). See Sec.~\ref{sec:Brownian} for details on the BD simulations. }
	\vspace{-0.5 cm}
	\label{fig:phasediagram_singlechain}
\end{figure}

Thus, this theory gives two possible equilibrium phases (see Fig.~\ref{fig:phasediagram_singlechain}). For $\alpha < \alpha_c$ the chain is extended in space ($\rho \simeq 0$) and has low magnetisation which corresponds to heterogeneous epigenetic marks ($\phi^2=0$): we dub this phase \emph{swollen-disordered} (SD). For values of $\alpha > \alpha_c \simeq 3.96$ the chain is collapsed ($\rho > 0$) and nearly uniformly coloured $(\phi^2 > 0)$ and hence we dub this phase \emph{compact-ordered} (CO). \dmi{Even though the free energy in Eq.~\eqref{eq:freeenergysingle} is obtained via several approximations, our theory is in excellent agreement with Brownian Dynamics simulations~\cite{Michieletto2016prx} and it unambiguously predicts the presence of a discontinuous jump of the order parameters $\rho$ and $\phi$ at the transition}.
% indicating a first order transition at the critical point which was  suggested in Refs.~\cite{Michieletto2016prx} only  on the basis of numerical simulations. 

As mentioned in the introduction, one of the most puzzling aspects of epigenetics is that, while it allows plastic transformations of cells or even organisms, it can be inherited even across generations. It is thus intriguing that our model predicts a first order transition for epigenetically-driven folding since its discontinuous nature provides a mechanism that endows memory to the system. For instance, once a chromosome has been taken over by a single epigenetic mark -- e.g., in the case of the inactive-X in female mammalian cells~\cite{Pinter2012,Nicodemi2007} -- our model predicts that such a state would be robust even against extensive perturbations such as those occurring during mitosis. Accordingly, it is well known that, while the initial choice of which of the two X's to inactivate is stochastic~\cite{Nicodemi2007}, this choice is remarkably robust across cell division; this balance between stochasticity and robustness of epigenetic processes can be appreciated everyday as it is manifested in the patchy coat of calico cats~\cite{Henikoff2016}. 

%%%%%%%%%%%%%%%% begin figure %%%%%%%%%%%%%%%%%%%
%%%%%%%%%%%%%%%% end figure %%%%%%%%%%%%%%%%%%% 

Our findings do not exclude, but rather complement, other mechanisms that are known to play important roles to ensure the correct transmission of a certain transcriptional programme, e.g.  non-coding nuclear RNA~\cite{Michieletto2019rnareview}, genomic bookmarks~\cite{Festuccia2016,Wang2013} or specialised transcription factors~\cite{Egli2008}.

Finally, we highlight that the two phases predicted by the equilibrium model do not capture in full the epigenetic-conformational states observed \emph{in vivo}. In the next section we provide a non-equilibrium generalisation of our model in order to obtain a wider spectrum of possible phases.

\section{Non-Equilibrium Model}
\label{sec:noneqmodel}
From the free energy in Eq.~\eqref{eq:freeenergysingle}, we can derive dynamical equations for $\rho$ and $\phi$. Since both order parameters are non-conserved  (here $\rho$ should be understood as the density of beads within the smallest box containing the polymer chain), we can obtain them by computing the steepest descent to the free energy minimum, i.e. $\partial_t \phi \sim - \delta \mathcal{H}/ \delta \phi$ and $\partial_t \rho \sim - \delta \mathcal{H}/ \delta \rho$, where $\mathcal{H}$ is the  energy functional $\mathcal{H}[\rho,\phi]=\int d\bm{x} \left[ f + \kappa_\phi (\nabla \phi)^2 + \kappa_{\rho} (\nabla \rho)^2 \right]$. This set of coupled equations is analogous to the equation defining ``Model A'' for the dynamics of non-conserved fields~\cite{Hohenberg1977,ChaikinLubensky}, and for our free energy they read
\begin{eqnarray}\label{crit_nondyn_eq_single}
\partial_t \phi ({\bf r},t) &=& \Gamma_{\phi} \left( -\frac{9\phi}{5\alpha_{\phi}\rho}-\frac{3}{e^{3\phi/2}+2} + \frac{6}{5} + \kappa_{\phi}\nabla^2 \phi \right ) ,  \\ 
\partial_t \rho ({\bf r},t) &=& \Gamma_{\rho} \left(\frac{9\phi^2}{10\alpha_{\rho}\rho^2}+ \frac{\alpha_{\rho}}{10}+\frac{\ln(1-\rho)}{\rho^2}+\frac{1}{\rho} + \kappa_{\rho}\nabla^2\rho\right )  \, , \notag
\end{eqnarray}
where $\Gamma_{\rho/\phi}$ and $\kappa_{\rho/\phi}$ are mobilities and surface tension-like coefficients, respectively. In Eqs.~\eqref{crit_nondyn_eq_single} we decouple $\alpha$ into two independent parameters affecting the dynamics of the polymer ($\alpha_\rho$) and of the epigenetic field ($\alpha_\phi$) separately. 
The case $\alpha_\rho \neq \alpha_\phi$ leads to non-equilibrium dynamics as these equations can no longer be derived from a free energy. 
Since the dynamics of chromatin and that of the epigenetic marks are very different processes there is no \emph{a priori} constraint for which $\alpha_{\phi}$ should be equal to $\alpha_{\rho}$. %are in principle allowed tune $\alpha_{\phi}$ and $\alpha_{\rho}$ independently. 
%Notice that a similar strategy has recently been employed to obtain an ``activity-driven'' phase separation in polymer melts by assigning different effective temperatures to different polymers~\cite{Smrek2017} (albeit there were no direct connection with epigenetic marks in that case). 

\begin{figure}[t!]
	\centering
	\includegraphics[width=0.48\textwidth]{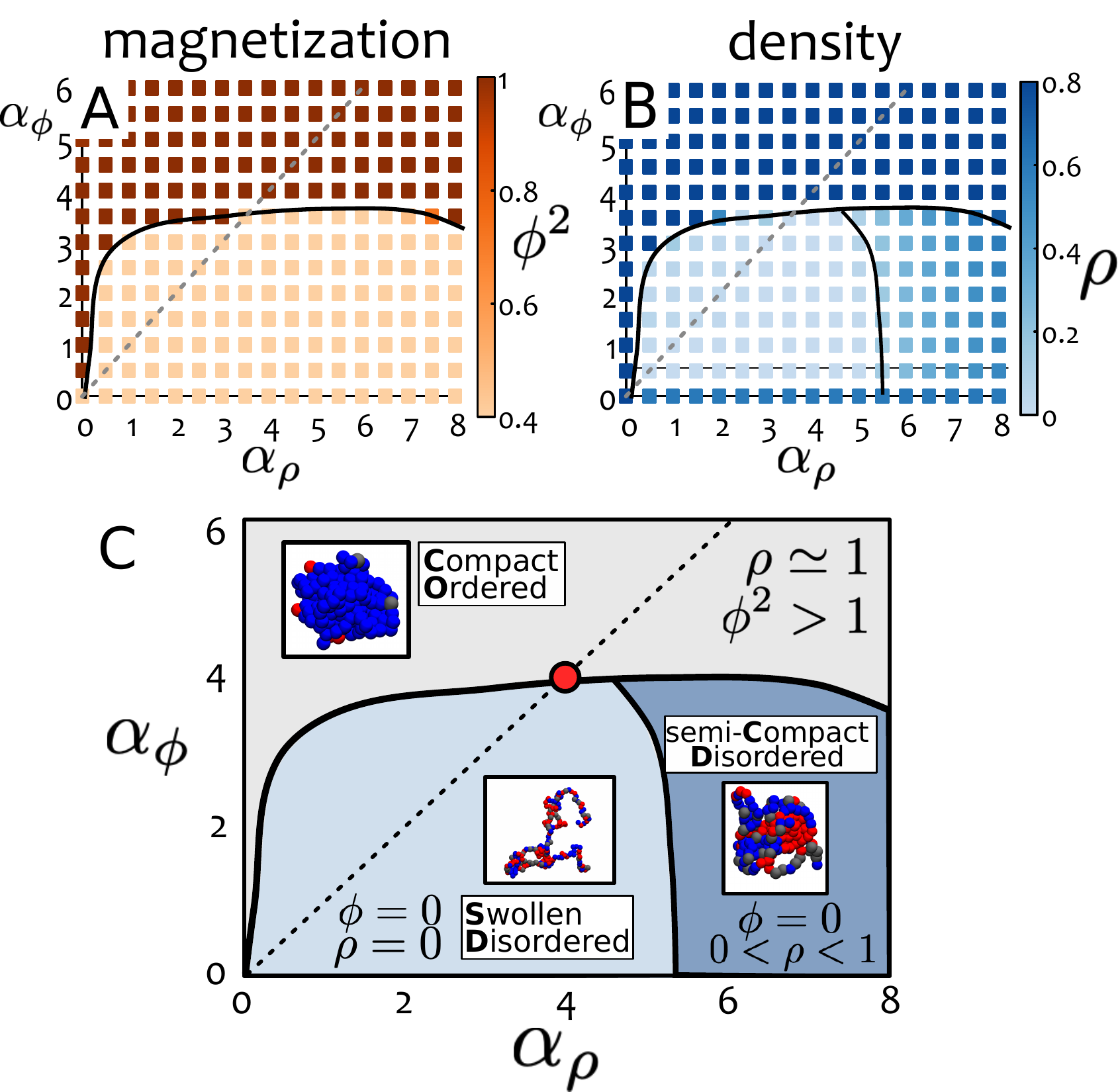}
	\caption{\textbf{Non-Equilibrium Phase Diagram for Single Chromosomes.} Steady state values of \textbf{A} magnetization and \textbf{B} density obtained by evolving Eqs.~\eqref{crit_nondyn_eq_single} on a $50\times 50$ grid with $\Gamma_{\phi}=0.1$, $\Gamma_{\rho}=0.001$ and $\kappa_\phi=\kappa_\rho=2$. \textbf{C} By cross-checking the two phase diagrams, we discover three regions with different combinations of $\rho$ and $\phi^2$. In addition to the Swollen-Disordered ($\rho=0$, $\phi^2=0$) and Compact-Ordered ($\rho=1$, $\phi^2>1$) phases found in equilibrium, we now observe a semi-Compact-Disordered (sCD) phase ($0<\rho<1$, $\phi^2=0$). The red circle denotes the first-order transition seen in equilibrium. (Inset) Snapshots of representative configurations from BD simulations of a magnetic polymer $N=200$ beads long (see Sec.~\ref{sec:Brownian} for details). Red and blue beads represent competing epigenetic states (or spins) whereas the grey ones are unmarked (neutral) states. 
	}
	%In clockwise direction from top left: CO=($T_L=0.9$,$T_R=0.9$), sCD=($T_L=0.7$,$T_R=2$), SD=($T_L=1.5$,$T_R=1.5$). Notice that $T_R=T_L$ implies that the system is in equilibrium as it obeys detailed balance. } 
	\vspace*{-0.3 cm}
	\label{fig:phasediagram_single_noneq}
\end{figure}

We have numerically integrated Eqs.~(\ref{crit_nondyn_eq_single}) using a standard Euler method with time step $dt=0.01$ on a $50 \times 50$ grid and setting $\Gamma_\phi=0.1$, $\Gamma_\rho=0.001$, $\kappa_\phi=\kappa_\rho=2$. 
The choice of these parameters only affect the kinetics of the process, not its long-time steady state. %but it should be mentioned that the right combination of timestep $dt$ and mobilities $\Gamma_\rho$ should be found to ensure numerical stability. 
The initial condition was set so that each lattice site had a random density and magnetisation broadly distributed around the means $\bar{\rho}=0.6$ and $\bar{m}=1$. We note that by choosing extreme values of density and magnetisation for the initial state may drive the equations to a local minimum and that Eqs.~\eqref{crit_nondyn_eq_single} include terms which diverge for $\rho \rightarrow 0$, and which are therefore prone to cause numerical blow-up in the swollen phase. Finally, by plotting the long-time steady state value of $\rho$ and $\phi$ for each combination of $(\alpha_\rho,\alpha_\phi)$ we can compile a non-equilibrium phase diagram which is shown in Fig.~\ref{fig:phasediagram_single_noneq}.  

Outside the equilibrium line $\alpha_{\rho}=\alpha_{\phi}$, a new phase characterized by 
a density greater than zero but a small total magnetisation ($0<\rho<1$, $\phi^2=0$) is observed. This phase transition is reminiscent of a collapsed conformation in which the epigenetic marks are heterogeneously distributed and we thus dub this phase semi-Compact-Disordered (CD).
%The latter is reminiscent of not collapsed conformations with homogeneous epigenetic marks and we dub this semi-Swollen-Disordered (CD).
Biologically, this phase may be similar to the one identified \emph{in vivo} with ``gene deserts'' in which no epigenetic mark is clearly dominating (also called ``black'' chromatin in Ref.~\cite{Sexton2012,Filion2010} where it was first discovered). These regions have been recently show to display compact conformations~\cite{Szabo2018} but the mechanisms behind such collapse are still unclear. Our theory suggests that even in absence of a uniform epigenetic pattern,  these regions may assume collapsed conformations in the limit in which the interaction between similar epigenetic marks is large enough to overcome the entropic penalty due to polymer folding.

%Biologically, this epigenetic phase may be similar to one that is seen in vivo:  for instance, chromatin regions with homogeneous epigenetic marks are typically associated with gene-rich regions when open (sSO)~\cite{Boettiger2016,Gilbert2004,Nozawa2017} whereas to silenced ``heterochromatin'' ones when collapsed (CO)~\cite{Boettiger2016,Falk2018}. On the other hand, epigenetically disordered regions, i.e. where no epigenetic mark is clearly dominating, can be seen in so-called ``gene deserts''~\cite{Sexton2012} and it is still unclear whether these regions assume collapsed or open conformations.

Through BD simulations we also observe a fourth phase that is not obtained through the model A equations (see Sec.~\ref{sec:Brownian} and Fig.~\ref{fig:bd_single}): one with uniform epigenetic coloring but swollen conformations. The lack of this phase in our theoretical phase diagram in Fig.~\ref{fig:phasediagram_single_noneq} may be due to the approximation of almost space-filling (Hamiltonian) walks that we used earlier to derive the free energy in Eq.~\eqref{eq:freeenergysingle}. Additionally, it should be noted that our field theory cannot fully resolve the chain structure 
of the polymer backbone and thus is not expected to sustain  a large magnetisation at low density. On the other hand, the polymer simulations presented in the next section fully account for chain topology; one may thus argue that this constraint 
is sufficient to create a local uniformity of epigenetic marks, even at low polymer density as in the swollen phase.

%These two phases were previously suggested by non-equilibrium Brownian Dynamics simulations~\cite{Michieletto2016prx,Michieletto2018nar}, and this 

\section{Brownian Dynamics Simulations of Magnetic Polymers} \label{sec:Brownian}

\subsection{Model}
\label{sec:Brownian_model}
To validate our findings within a more realistic polymer framework, we perform Brownian Dynamics (BD) simulations of chromosomes with dynamic epigenetic marks. 
We model chromosomes as semi-flexible chains with beads of size $\sigma$~\cite{Rosa2008} and mark each bead with a ``spin'' or epigenetic state $q=\{-1,0,1\}$. Such a co-polymer model for chromosomes has been successfully employed in the literature~\cite{Brackley2016nar,Jost2014B}. We simulate the dynamics of each of the polymer segments via a Langevin equation at the temperature $T_L$, i.e. 
\begin{equation}
 m \dfrac{d^2 \bm{r}}{dt^2} = - \nabla U - \gamma \dfrac{d \bm{r}}{dt} + \sqrt{2 k_BT_L} \bm{\eta}(t),
\end{equation}
where $\bm{r}$ is the position of the bead, $\gamma$ describes the bead friction and $\bm{\eta}$ is a noise satisfying the fluctuation-dissipation theorem~\cite{Kremer1990}.
The potential $U$ is a contribution of three force fields which model steric interactions, chain connectivity and rigidity. More specifically we consider $U =\mathrm{U}_{\text{LJ}} + \mathrm{U}_{\text{H}}  + \mathrm{U}_{\text{K}}$, where
\begin{equation}\label{eq:Kratky_Potts}
\mathrm{U}_{\text{K}}  =  \frac{k_B T_L \ell_P}{\sigma}\sum_{i=1}^{M-2}\left( 1 - \frac{ \vec{u}_i \cdot \vec{u}_{i+1}}{\norm{\vec{u}_i}\norm{\vec{u}_{i+1}}}\right) \text{ ,}
\end{equation}
is the bending potential with $\vec{u}_j \equiv \vec{r}_{j+1}-\vec{r}_j$. In this equation, $\ell_P$ is identified with the persistence length of the chain, here set to $\ell_P = 3 \sigma \simeq 90$ nm to roughly match that of chromatin~\cite{Socol2019}.
Then to model the connectivity of the chain we set springs between consecutive beads 
\begin{equation}\label{eq:Harmonic_Potts}
U_{\text{H}} = \sum_{i=1}^{M-1} \frac{k_h}{2} \left( \left| \vec{r}_i - \vec{r}_{i+1} \right| - r_0\right)^2 \, ,
\end{equation}
where $k_h$ is a spring  constant set to $k_H = 200 \epsilon = 200 k_B T_L$ and $r_0 = 1.1 \sigma$. 
The key force field is the Lennard-Jones potential which modulates the repulsive/attractive interaction between beads. Importantly, we choose this pairwise force field to depend on the spins of the interacting pair as follows:
\begin{equation} \label{eq:LJ}
	U_{\text{LJ}}(r;q_i, q_j) = \frac{ 4  \varepsilon(q_i,q_j) }{\mathcal{N}} \left[ \left(\frac{\sigma}{r}\right)^{12}
	-  \left(\frac{\sigma}{r}\right)^6 - U_0(r_c) \right] \, , 
\end{equation}
for $r< r_c(q_i,q_j)$ and $0$ otherwise; in this equation, $U_0$ is an auxiliary function which ensures that $U_{\text{LJ}}(r_c; q_i, q_j) = 0$ and $\mathcal N$ is a normalisation so that the minimum of the potential is at $-\varepsilon$.
 
The essential parameter here is the cutoff $r_c(q_i,q_j)$ which is $q-$dependent and set so that beads with the same spin value, or epigenetic mark, attract one another whereas beads with different marks (i.e., different $q$)  interact only repulsively. 
This is done by choosing:
\begin{equation}
r_c(q_i, q_j) = 
\begin{cases}
 2^{1/6}\sigma \text{ if } q_i \neq q_j \text{ or } q_i=q_j = 0 \\
1.8\sigma \text{ if } q_i = q_j \neq 0 \, .\\
\end{cases}
\end{equation}
With these choices the pair interaction between equally colored (but not neutral) beads displays a dip which models short range attraction, whereas it is fully repulsive (no dip) for differently colored (or neutral) beads. Finally, the free parameter $\varepsilon(q_i, q_j)$ is set to $\varepsilon$ if $q_j=q_j \neq 0$ and 1 otherwise. These choices mimic a ferromagnetic interaction (with strength $\varepsilon$) which tend to align marks that are nearby 
or to attract beads with similar marks.

We then evolve the equations of motion for each bead in the system using fixed-volume and constant-temperature Brownian dynamics (BD) simulations (NVT ensemble). The simulations are run within the LAMMPS engine~\cite{Plimpton1995} and the equations of motion are integrated using a velocity Verlet algorithm, in which all beads are weakly coupled to a heat bath with friction $\gamma = \tau_B^{-1}$ where $\tau_B = 3\pi \eta \sigma^3/k_BT$ is the self-diffusion (Brownian) time of a bead of size $\sigma$ moving in a solution with viscosity $\eta$. % (which we consider to be water, i.e. $\eta= 1 cP$, for the mapping to real units). 
Finally, the integration time step is set to $\Delta \tau = 0.01 \, \tau_B$. The polymer dynamics subject to thermal fluctuations is then interleaved with the recolouring dynamics of the polymer beads which we now describe.

\begin{figure}[t!]
	\centering
	\includegraphics[width=0.4\textwidth]{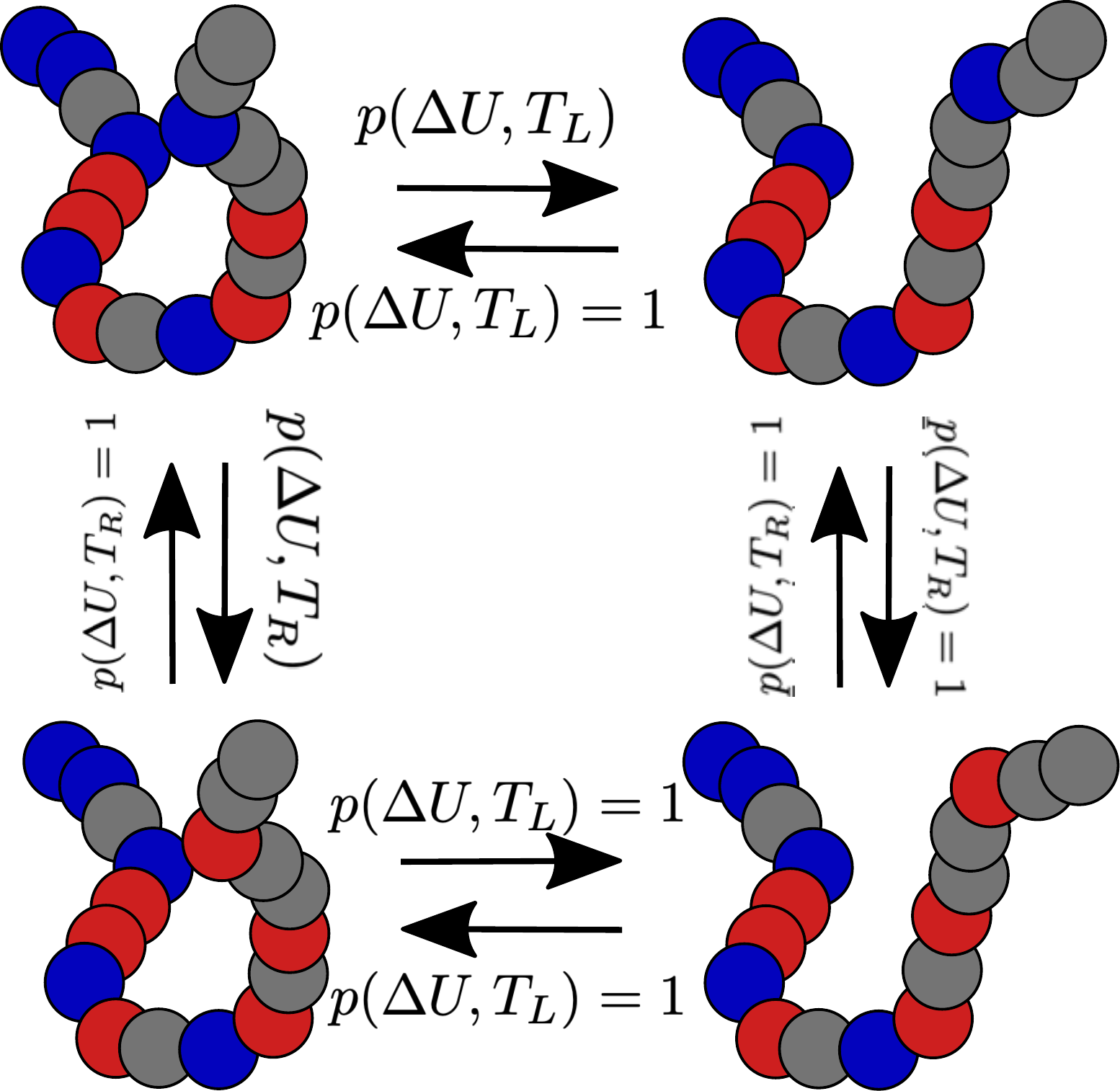}
	\caption{\textbf{Detailed Balance is Broken when $T_L\neq T_R$.} This figure shows a Kolmogorov loop over some states of the polymer. The product of the transition probabilities in the clockwise loop gives $p(\Delta U,T_L)$ whereas the counter-clockwise loop gives $p(\Delta U,T_R)$. The Kolmogorov criterion for obeying detailed balance is that the two must be equal, which is true only when $T_L=T_R$.  } 
	\label{fig:kolmogorov_loop}
\end{figure}

Each bead along the chain can change its epigenetic state, or spin, at rate $\kappa_s =\tau_R^{-1}$, or in other words, on average every $\tau_R$ steps each bead is picked once and an attemp is made to change its color into a different one. %. In each of these attempts, we randomly select a bead and randomly change its colour to a different one. 
If the move lowers the energy of the system we accept it, otherwise we assign an acceptance probability $p=e^{-\Delta U/k_BT_R}$ where $\Delta U$ is the difference between the system energy after and before the move.

\begin{figure*}[t!]
	\centering
	\includegraphics[width=0.95\textwidth]{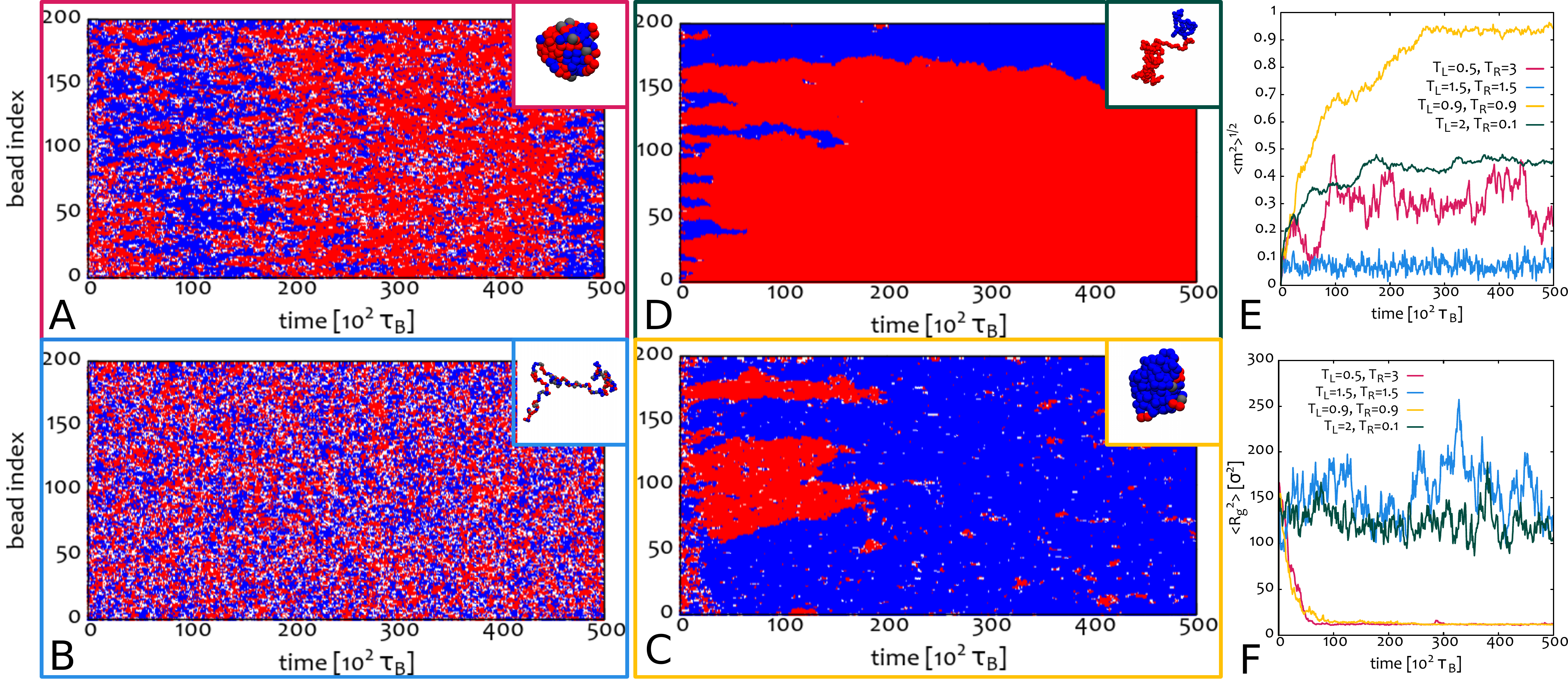}
	\caption{\textbf{BD simulations of single chromosomes out-of-equilibrium.} \textbf{A-D}  Kymographs showing the evolution of the epigenetic marks along the polymer in time (insets show typical conformations). In counterclockwise order: \textbf{A} $T_L=0.5,T_R=3$; \textbf{B} $T_L=1.5,T_R=1.5$; \textbf{C} $T_L=0.9,T_R=0.9$; \textbf{D} $T_L=2,T_R=0.1$. All temperatures are in units of the room temperature $T^*=300K$. \textbf{E-F} Time evolution of  magnetisation $\langle m^2 \rangle^{1/2}$ (\textbf{E}) and radius of gyration $\langle R_g^2 \rangle$ (\textbf{F}) averaged over several independent simulations. Note the, in the long time limit, both observables reach a well defined steady state whose value depend on the pair $(T_L,T_R)$. The boxes around (\textbf{A-D}) match the color scheme in panels \textbf{E-F}.  } 
	\label{fig:bd_single}
\end{figure*}

It is important to notice that the dynamics of the chain is subject to thermal fluctuations controlled by the temperature $T_L$, whereas the Metropolis algorithm on the beads recolouring is weighted by an effective temperature $T_R$ with, \emph{a priori}, $T_L \neq T_R$. 
One can show that this condition breaks detailed balance by constructing a Kolmogorov loop over some of the states of the system as shown in Fig.~\ref{fig:kolmogorov_loop}: the product of the transitions in the clockwise direction is equal to $p(\Delta U, T_R)= e^{-\Delta U/k_BT_R}$ whereas the one over the counter-clockwise loop to $p(\Delta U, T_L)= e^{-\Delta U/k_BT_L}$. The Kolmogorov criterion states that detailed balance is obeyed only if the two are equal, which is true only if $T_L=T_R$. 

As mentioned above, there is no biological constraint for which $T_L$ should be equal to $T_R$; in fact, chromatin and epigenetic mark dynamics are very different biophysical processes. In our model, this is captured by the fact that their effective temperatures are independently tuned. 

Another possible way to violate detailed balance in this model is by introducing a fourth bead type that cannot be magnetised (we call this the ``off'' state as opposed to the standard ``on'' state). By randomly switching off beads with any $q$ and activating ``off'' beads to one particular state (say $q=0$), one can set up a current in the epigenetic states and therefore break time-reversal symmetry (and detailed balance). %This switching strategy sets up a circular current in epigenetic states and drives the system out-of-equilibrium. 
Biologically, this switching between on and off states may correspond to chromatin regions whose assembly in nucleosomes is transiently disrupted and cannot bear epigenetic marks or to proteins that can switch between different conformations due to ATP-binding and hydrolysis or phosphorylation~\cite{Brackley2017biophysj,Michieletto2019rnareview} and it is explored in more detail in the companion paper~\cite{PRL}. 

Finally, it should be noted that the correspondence between PDEs (such as the ones derived and numerically solved above) and BD is qualitative and we do not infer BD parameters from PDEs or viceversa. For instance, the proposed PDEs do not take into account polymer backbone, self-avoidance and noise (which are instead fully accounted for in BD simulations). On the other hand, the measured observables are the same: local concentration of chromatin and magnetisation, can be considered to define different phases and compile a phase diagram for BD simulations (see below) which can be qualitatively compared with the results from PDEs (see also Ref.~\cite{Michieletto2018nar}). 

\subsection{Results for Isolated Magnetic Polymers}
\label{sec:Brownian_single_chrom}

In order to verify that the theories obtained in Eq.~\eqref{eq:freeenergysingle} and Eqs.~\eqref{crit_nondyn_eq_single} predict behaviours that are in agreement with more refined models accounting for chain connectivity and self-avoidance constraints, we first perform Brownian Dynamics simulations of the magnetic polymer model described in the previous section considering dilute conditions -- i.e., one chain in free space. 

We always initialise the system from a state in which a polymer is in a self-avoiding walk conformation and with random epigenetic marks. To obtain stable states of the system we monitor the evolution of the radius of gyration  $R_g$ of an $N$-beads long chain, defined as 
\begin{equation}
	R_g^2 = \dfrac{1}{N^2}\sum_{i=1}^N \sum_{j=1}^N \left|\bm{r}_i - \bm{r}_j \right|^2 
\end{equation}
and of its epigenetic magnetisation $m$, defined as 
\begin{equation}
	m = \dfrac{n(1) - n(-1)}{N} \, ,
\end{equation}
where $n(q)$ is the number of beads with epigenetic state $q$. Finally, we assign a state based on the steady values attained by these two observables, which can be respectively related to $\rho$ and $\phi^2$ used in the theories above. 
Examples of BD simulations of single magnetic polymers are reported in Fig.~\ref{fig:bd_single} where kymographs show the evolution of $q$ along the chain index as a function of time. We also report typical conformations and average over different replicas. % of magnetisation and gyrations radius plotted against time. 

\begin{figure*}[t!]
	\centering
	\includegraphics[width=0.95\textwidth]{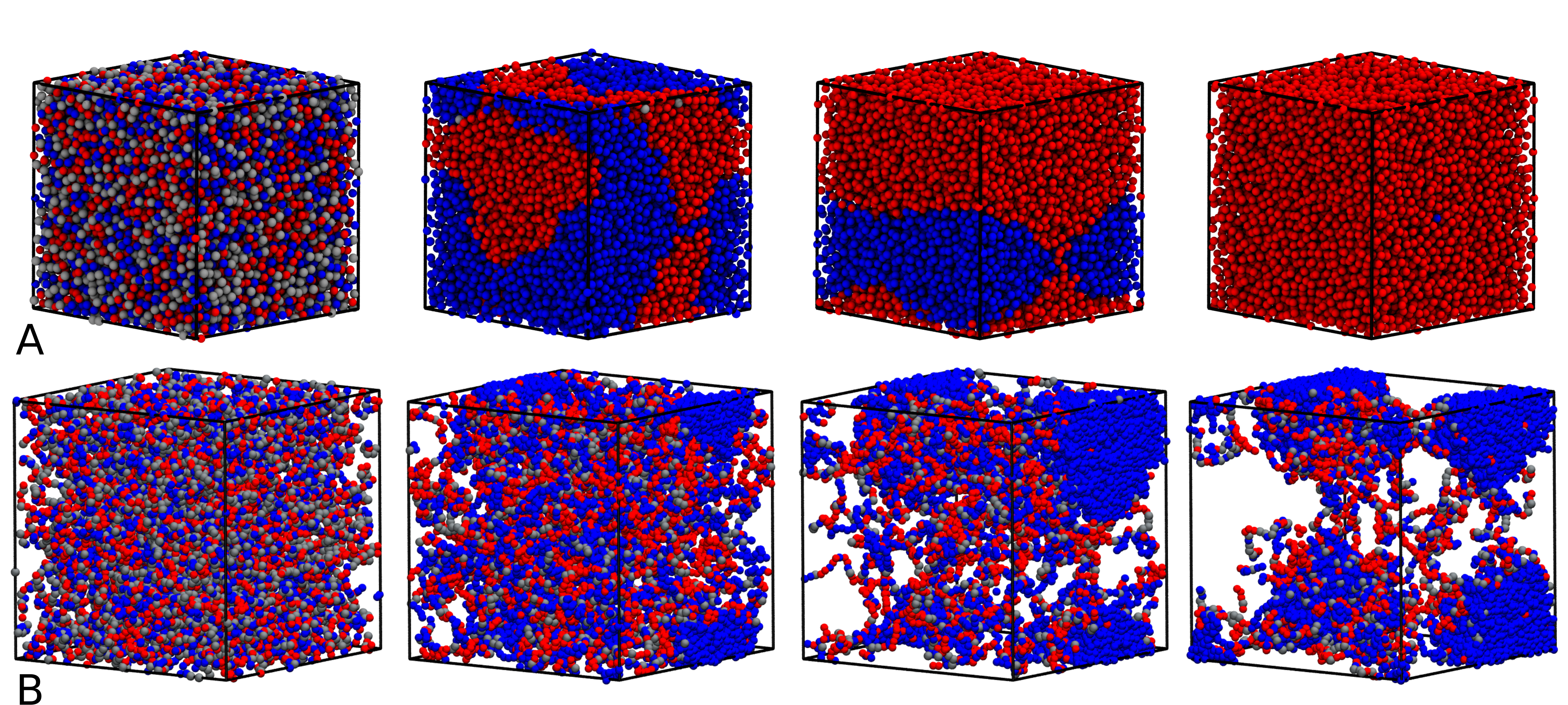}
	\caption{\textbf{BD simulations of magnetic Polymer Melts.} Time evolution  of a magnetic polymer melt with $N=50$ chains each $M=256$ beads long. Panel \textbf{A} refers to a system with monomer density $\rho= NM/V = 0.8\sigma^{-3}$ and attraction strength $\varepsilon= k_BT_L$, while panel \textbf{B} to one with  $\rho= 0.1\sigma^{-3}$ and $\varepsilon= 0.86 k_BT_L$. In both cases $T_R=T_L$. } 
	\label{fig:melt}
\end{figure*}

For the equilibrium case (Fig.~\ref{fig:phasediagram_singlechain}) the key parameter is the depth of the Lennard-Jones attraction between equal non-neutral marks which is set to $\varepsilon$. As shown in Figure~\ref{fig:phasediagram_singlechain} and Fig.~\ref{fig:bd_single}(B-C), as we vary $\varepsilon$ (with $T_L=T_R=1$ in units of the room temperature $T^*=300K$) we find phases that are matching the ones expected from the free energy in Eq.~\eqref{eq:freeenergysingle}, separated by a first order transition at the critical point $\varepsilon_c \simeq 0.9$~\cite{Michieletto2016prx}. 
It should be noted that these findings (and the ones below) are robust with respect to the choice of recolouring rate $\tau_R^{-1}$ and initial conditions.

To verify the non-equilibrium case reported in Fig.~\ref{fig:phasediagram_single_noneq}, we fix $\varepsilon=1$ and tune $T_L$ and $T_R$ independently. As mentioned above, we find the compact-disordered phase predicted by the theory but we also find a swollen-ordered phase as shown in Fig.~\ref{fig:bd_single} and Refs.~\cite{Michieletto2016prx,Michieletto2018nar}. We argue that this is because, as mentioned above, Eqs.~\eqref{crit_nondyn_eq_single} are prone to numerical instabilities in the regime $\rho \rightarrow 0$ and are obtained assuming a space-filling curve (Hamiltonian walk). 

Nevertheless, it is remarkable to notice how robust and generic our findings are, as we obtain very similar conformations in both the analytical model and BD simulations. 
Deriving a precise mapping between the parameters of our theory, such as $\alpha_{\phi}$ and $\alpha_{\rho}$ and the ones of the BD simulations such as $T_L$ and $T_R$ or $\epsilon$ is complicated, as the underlying processes are very different. For instance, the former is based on continuum equations, whereas the latter on the diffusion of discrete beads that are linked together on a chain. Moreover, the Time-Dependent Landau-Ginzburg equations above are solved at zero temperature whereas the BD simulations are run through a Langevin equation accounting for thermal fluctuations and the recolouring dynamics through a Monte-Carlo scheme. In spite of all this, we observe similar regimes thus suggesting that the uncovered physics is not system-dependent and therefore universal. 

\subsection{Results for Magnetic Polymer Melts}
\label{sec:Brownian_melts}

In our companion paper~\cite{PRL} we present a theory for how epigenetic marks can drive the compartmentalisation of the genome in the nucleus. This theory can also be checked with more refined BD simulations of magnetic polymers. The difference with respect to the set up described above is that now the genome in the cell nucleus should be modelled as a melt of, rather than isolated, magnetic polymers. 

To do this we prepare linear chains as random walks in a box with periodic boundary conditions and reduce the box size until the desired monomer density $\rho=NM/L^3$ is attained. We typically consider $N=50$ polymers with $M=256$ beads each and the range of parameters employed are $\rho=0.1$ -- $0.8$ $\sigma^{-3}$ and $\varepsilon/k_BT_L=0.75$--$1.1$. Snapshots of this set up and its evolution in time for two choices of $\rho$ and $\varepsilon$ are reported in Fig.~\ref{fig:melt}. In the first case ($\rho=0.8\sigma^{-3}$, $\varepsilon=k_BT_L$) we observe bicontinuous spanning clusters that merge into one without instabilities in the density, as predicted by our mean-field theory (see Ref.~\cite{PRL}) for quenches outside the coexistence region. In the second, ($\rho=0.1\sigma^{-3}$, $\varepsilon=0.86k_BT_L$) we observe a phase separation of the system in dense epigenetically ordered regions surrounded by unoccupied/sparse regions again in line with our theory for quenches within the coexistence region. It should be noted that both cases are in equilibrium as $T_L=T_R$. More examples are given in the companion paper~\cite{PRL}.

This magnetic polymer melt can be extended to non-equilibrium by adding a switching process as described above. Specifically, we can turn off (or inactivate) a bead at rate $\sigma_i$ and re-activate any ``off'' bead at rate $\sigma_a$ by changing its state to neutral ($q=0$). This choice sets up a current in the epigenetic states which causes the system to violate details balance and time-reversal symmetry (see Ref.~\cite{PRL}). We have qualitatively explored this system for selected parameter choices in~\cite{PRL}. A more systematic analysis would require large-scale computer simulations and will be pursued elsewhere. 

\section{Conclusions}
\label{sec:conclusions}
In this work we have proposed a field theoretical approach to study the interplay between the dynamics of epigenetic marks along chromosomes \emph{in vivo} and their spatial 3D organization. By mapping the system onto a model of magnetic polymers whose spin variables describe the epigenetic marks, we have analytically established that the transition between the swollen epigenetically disordered phase and the compact epigenetically coherent one is first order. The discontinuous nature of the transition, confirmed by Brownian dynamics simulations of a more realistic model of magnetic polymers, is a genuine product of the competition between the collective dynamics of epigenetic marks and the chain conformation and provides a mechanism of bistable epigenetic switch which endows the system with memory.
\dmi{It is interesting to note that in our companion paper~\cite{PRL}, where we study a model for the full nucleus, we obtain instead  a continuous transition between uniform disordered and uniform ordered phases. We argue that this is due to the fact that, while the model analysed here for single chromosomes is globally non-conserved in both density and magnetisation, the model for the full (interphase) nucleus has to conserve the overall density of DNA. The biological implications for this qualitative difference is that there may be additional ingredients needed to retain memory of epigenetic and conformational states in a nucleus which is homogeneously filled with chromatin.} %uniformly in  might depend on chromatin densityassumed by single chromosomes have a window of metastability which might render them more robust against external perturbations with respect to the overall nuclear organisation. }
%Therefore, while the whole genome can switch between disordered and ordered phases, some of the chromosomes may retain their epigenetic state. This may be particularly relevant for the inactive X-chromosome, which is robustly remembered across cell division while the overall genome loses its epigenetic marks~\cite{Wang2013}. }

By allowing the time-dependent Ginzburg-Landau equations of the model to follow non-equilibrium dynamics, we also found a third (non-equilibrium) phase. This is characterized by compact states in which the epigenetic marks are incoherently distributed and is reminiscent of gene deserts (or ``black'' chromatin~\cite{Filion2010}) observed in \emph{Drosophila}. Finally, by using the corresponding BD simulations (with broken detailed balance) we have been able to perform a wider exploration of the non-equilibrium phase diagram and to observe the predicted compact disordered phase as well as a new phase characterized by extended chain conformations that are coherently coloured.

These results and the ones obtained with similar techniques but at the scale of the full nucleus (see ~\cite{PRL}) show how statistical mechanics and field theory approaches on models of magnetic polymers can contribute to pinpoint the general multiscale physical mechanisms that govern the interplay between epigenetic spreading and genomic organization in the nucleus. \dmi{In particular, our strategy allows us to weakly push the system out of equilibrium and to study this problem by using a (semi-)analytical framework (Figs.~\ref{fig:f_Single}-\ref{fig:phasediagram_single_noneq}); this would be impossible (or a lot harder) to do in a far-from-equilibrium model without an underlying free energy.} % On the other hand, we should note that our generic dynamical equations cannot unambiguously reflect any particular biophysical process.}
\dmi{In the future, it would be interesting to explore other models tailored to capture specific non-equilibrium processes} such as the disruption of chromatin structure by polymerase, or remodelling during mitosis. The looping mediated by slip-link-like proteins such as cohesins and condensins might also be an important ingredient to include~\cite{Orlandini2019,Fudenberg2016,Brackley2017prl} while the introduction of quenched spins that seed domain formation -- as in genomic bookmarking~\cite{Michieletto2018nar} -- can be easily incorporated at a field-theoretic level.

We hope that technical advances will soon make it possible to perform experiments with epigenetic marks on reconstituted chromatin \emph{in vitro}~\cite{Tauran2019}; this should give more hints on how to extend polymer magnetic models to improve our understanding of both epigenetic spreading processes~\cite{Michieletto2017scirep,Tauran2019} and the generic interplay between different read-write protein complexes. 

\section*{Acknowledgements}
We thank the European Research Council (ERC CoG 648050 THREEDCELLPHYSICS) for funding.  DMi and EO would also like to acknowledge the networking support by EUTOPIA (CA17139).

\bibliographystyle{apsrev4-1}
\bibliography{library}

\begin{thebibliography}{53}%
\makeatletter
\providecommand \@ifxundefined [1]{%
 \@ifx{#1\undefined}
}%
\providecommand \@ifnum [1]{%
 \ifnum #1\expandafter \@firstoftwo
 \else \expandafter \@secondoftwo
 \fi
}%
\providecommand \@ifx [1]{%
 \ifx #1\expandafter \@firstoftwo
 \else \expandafter \@secondoftwo
 \fi
}%
\providecommand \natexlab [1]{#1}%
\providecommand \enquote  [1]{``#1''}%
\providecommand \bibnamefont  [1]{#1}%
\providecommand \bibfnamefont [1]{#1}%
\providecommand \citenamefont [1]{#1}%
\providecommand \href@noop [0]{\@secondoftwo}%
\providecommand \href [0]{\begingroup \@sanitize@url \@href}%
\providecommand \@href[1]{\@@startlink{#1}\@@href}%
\providecommand \@@href[1]{\endgroup#1\@@endlink}%
\providecommand \@sanitize@url [0]{\catcode `\\12\catcode `\$12\catcode
  `\&12\catcode `\#12\catcode `\^12\catcode `\_12\catcode `\%12\relax}%
\providecommand \@@startlink[1]{}%
\providecommand \@@endlink[0]{}%
\providecommand \url  [0]{\begingroup\@sanitize@url \@url }%
\providecommand \@url [1]{\endgroup\@href {#1}{\urlprefix }}%
\providecommand \urlprefix  [0]{URL }%
\providecommand \Eprint [0]{\href }%
\providecommand \doibase [0]{http://dx.doi.org/}%
\providecommand \selectlanguage [0]{\@gobble}%
\providecommand \bibinfo  [0]{\@secondoftwo}%
\providecommand \bibfield  [0]{\@secondoftwo}%
\providecommand \translation [1]{[#1]}%
\providecommand \BibitemOpen [0]{}%
\providecommand \bibitemStop [0]{}%
\providecommand \bibitemNoStop [0]{.\EOS\space}%
\providecommand \EOS [0]{\spacefactor3000\relax}%
\providecommand \BibitemShut  [1]{\csname bibitem#1\endcsname}%
\let\auto@bib@innerbib\@empty
%</preamble>
\bibitem [{\citenamefont {Waddington}(1942)}]{Waddington1942a}%
  \BibitemOpen
  \bibfield  {author} {\bibinfo {author} {\bibfnamefont {C.~H.}\ \bibnamefont
  {Waddington}},\ }\href {\doibase 10.1038/150563a0} {\bibfield  {journal}
  {\bibinfo  {journal} {Nature}\ }\textbf {\bibinfo {volume} {150}},\ \bibinfo
  {pages} {563} (\bibinfo {year} {1942})}\BibitemShut {NoStop}%
\bibitem [{\citenamefont {Alberts}\ \emph {et~al.}(2014)\citenamefont
  {Alberts}, \citenamefont {Johnson}, \citenamefont {Lewis}, \citenamefont
  {Morgan},\ and\ \citenamefont {Raff}}]{Alberts2014}%
  \BibitemOpen
  \bibfield  {author} {\bibinfo {author} {\bibfnamefont {B.}~\bibnamefont
  {Alberts}}, \bibinfo {author} {\bibfnamefont {A.}~\bibnamefont {Johnson}},
  \bibinfo {author} {\bibfnamefont {J.}~\bibnamefont {Lewis}}, \bibinfo
  {author} {\bibfnamefont {D.}~\bibnamefont {Morgan}}, \ and\ \bibinfo {author}
  {\bibfnamefont {M.}~\bibnamefont {Raff}},\ }\href
  {http://books.google.com/books?id=1ZUDoQEACAAJ{\&}pgis=1} {\emph {\bibinfo
  {title} {{Molecular Biology of the Cell}}}}\ (\bibinfo  {publisher} {Taylor
  {\&} Francis},\ \bibinfo {year} {2014})\ p.\ \bibinfo {pages}
  {1464}\BibitemShut {NoStop}%
\bibitem [{\citenamefont {Cortini}\ \emph {et~al.}(2015)\citenamefont
  {Cortini}, \citenamefont {Barbi}, \citenamefont {Care}, \citenamefont
  {Lavelle}, \citenamefont {Lesne}, \citenamefont {Mozziconacci},\ and\
  \citenamefont {Victor}}]{Cortini2015}%
  \BibitemOpen
  \bibfield  {author} {\bibinfo {author} {\bibfnamefont {R.}~\bibnamefont
  {Cortini}}, \bibinfo {author} {\bibfnamefont {M.}~\bibnamefont {Barbi}},
  \bibinfo {author} {\bibfnamefont {B.~R.}\ \bibnamefont {Care}}, \bibinfo
  {author} {\bibfnamefont {C.}~\bibnamefont {Lavelle}}, \bibinfo {author}
  {\bibfnamefont {A.}~\bibnamefont {Lesne}}, \bibinfo {author} {\bibfnamefont
  {J.}~\bibnamefont {Mozziconacci}}, \ and\ \bibinfo {author} {\bibfnamefont
  {J.-M.}\ \bibnamefont {Victor}},\ }\href {\doibase 10.1007/JHEP10(2015)034}
  {\bibfield  {journal} {\bibinfo  {journal} {Rev. Mod. Phys.}\ }\textbf
  {\bibinfo {volume} {88}},\ \bibinfo {pages} {1} (\bibinfo {year}
  {2015})}\BibitemShut {NoStop}%
\bibitem [{\citenamefont {Tang}\ \emph {et~al.}(2015)\citenamefont {Tang},
  \citenamefont {Dietmann}, \citenamefont {Irie}, \citenamefont {Leitch},
  \citenamefont {Floros}, \citenamefont {Bradshaw}, \citenamefont {Hackett},
  \citenamefont {Chinnery},\ and\ \citenamefont {Surani}}]{Tang2015a}%
  \BibitemOpen
  \bibfield  {author} {\bibinfo {author} {\bibfnamefont {W.~W.~C.}\
  \bibnamefont {Tang}}, \bibinfo {author} {\bibfnamefont {S.}~\bibnamefont
  {Dietmann}}, \bibinfo {author} {\bibfnamefont {N.}~\bibnamefont {Irie}},
  \bibinfo {author} {\bibfnamefont {H.~G.}\ \bibnamefont {Leitch}}, \bibinfo
  {author} {\bibfnamefont {V.~I.}\ \bibnamefont {Floros}}, \bibinfo {author}
  {\bibfnamefont {C.~R.}\ \bibnamefont {Bradshaw}}, \bibinfo {author}
  {\bibfnamefont {J.~A.}\ \bibnamefont {Hackett}}, \bibinfo {author}
  {\bibfnamefont {P.~F.}\ \bibnamefont {Chinnery}}, \ and\ \bibinfo {author}
  {\bibfnamefont {M.~A.}\ \bibnamefont {Surani}},\ }\href {\doibase
  10.1016/j.cell.2015.04.053} {\bibfield  {journal} {\bibinfo  {journal}
  {Cell}\ }\textbf {\bibinfo {volume} {161}},\ \bibinfo {pages} {1453}
  (\bibinfo {year} {2015})}\BibitemShut {NoStop}%
\bibitem [{\citenamefont {Rulands}\ \emph {et~al.}(2018)\citenamefont
  {Rulands}, \citenamefont {Lee}, \citenamefont {Clark}, \citenamefont
  {Stegle}, \citenamefont {Simons},\ and\ \citenamefont
  {Correspondence}}]{Rulands2018}%
  \BibitemOpen
  \bibfield  {author} {\bibinfo {author} {\bibfnamefont {S.}~\bibnamefont
  {Rulands}}, \bibinfo {author} {\bibfnamefont {H.~J.}\ \bibnamefont {Lee}},
  \bibinfo {author} {\bibfnamefont {S.~J.}\ \bibnamefont {Clark}}, \bibinfo
  {author} {\bibfnamefont {O.}~\bibnamefont {Stegle}}, \bibinfo {author}
  {\bibfnamefont {B.~D.}\ \bibnamefont {Simons}}, \ and\ \bibinfo {author}
  {\bibfnamefont {W.~R.}\ \bibnamefont {Correspondence}},\ }\href {\doibase
  10.1016/j.cels.2018.06.012} {\ ,\ \bibinfo {pages} {63} (\bibinfo {year}
  {2018})}\BibitemShut {NoStop}%
\bibitem [{\citenamefont {Crews}(2003)}]{Crews2003}%
  \BibitemOpen
  \bibfield  {author} {\bibinfo {author} {\bibfnamefont {D.}~\bibnamefont
  {Crews}},\ }\href {\doibase 10.1046/j.1525-142X.2003.03008.x} {\bibfield
  {journal} {\bibinfo  {journal} {Evolution and Development}\ }\textbf
  {\bibinfo {volume} {5}},\ \bibinfo {pages} {50} (\bibinfo {year}
  {2003})}\BibitemShut {NoStop}%
\bibitem [{\citenamefont {Ernst}\ \emph {et~al.}(2015)\citenamefont {Ernst},
  \citenamefont {{Van Hiel}}, \citenamefont {Depuydt}, \citenamefont {Boerjan},
  \citenamefont {{De Loof}},\ and\ \citenamefont {Schoofs}}]{Ernst2015a}%
  \BibitemOpen
  \bibfield  {author} {\bibinfo {author} {\bibfnamefont {U.~R.}\ \bibnamefont
  {Ernst}}, \bibinfo {author} {\bibfnamefont {M.~B.}\ \bibnamefont {{Van
  Hiel}}}, \bibinfo {author} {\bibfnamefont {G.}~\bibnamefont {Depuydt}},
  \bibinfo {author} {\bibfnamefont {B.}~\bibnamefont {Boerjan}}, \bibinfo
  {author} {\bibfnamefont {A.}~\bibnamefont {{De Loof}}}, \ and\ \bibinfo
  {author} {\bibfnamefont {L.}~\bibnamefont {Schoofs}},\ }\href {\doibase
  10.1242/jeb.107078} {\bibfield  {journal} {\bibinfo  {journal} {Journal of
  Experimental Biology}\ }\textbf {\bibinfo {volume} {218}},\ \bibinfo {pages}
  {88} (\bibinfo {year} {2015})}\BibitemShut {NoStop}%
\bibitem [{\citenamefont {Grewal}\ \emph {et~al.}(2011)\citenamefont {Grewal},
  \citenamefont {Grewal},\ and\ \citenamefont {Moazed}}]{Grewal2011}%
  \BibitemOpen
  \bibfield  {author} {\bibinfo {author} {\bibfnamefont {S.~I.~S.}\
  \bibnamefont {Grewal}}, \bibinfo {author} {\bibfnamefont {S.~I.~S.}\
  \bibnamefont {Grewal}}, \ and\ \bibinfo {author} {\bibfnamefont
  {D.}~\bibnamefont {Moazed}},\ }\href {\doibase 10.1126/science.1086887}
  {\bibfield  {journal} {\bibinfo  {journal} {Science}\ }\textbf {\bibinfo
  {volume} {798}},\ \bibinfo {pages} {798} (\bibinfo {year}
  {2011})}\BibitemShut {NoStop}%
\bibitem [{\citenamefont {Boettiger}\ \emph {et~al.}(2016)\citenamefont
  {Boettiger}, \citenamefont {Bintu}, \citenamefont {Moffitt}, \citenamefont
  {Wang}, \citenamefont {Beliveau}, \citenamefont {Fudenberg}, \citenamefont
  {Imakaev}, \citenamefont {Mirny}, \citenamefont {Wu},\ and\ \citenamefont
  {Zhuang}}]{Boettiger2016}%
  \BibitemOpen
  \bibfield  {author} {\bibinfo {author} {\bibfnamefont {A.~N.}\ \bibnamefont
  {Boettiger}}, \bibinfo {author} {\bibfnamefont {B.}~\bibnamefont {Bintu}},
  \bibinfo {author} {\bibfnamefont {J.~R.}\ \bibnamefont {Moffitt}}, \bibinfo
  {author} {\bibfnamefont {S.}~\bibnamefont {Wang}}, \bibinfo {author}
  {\bibfnamefont {B.~J.}\ \bibnamefont {Beliveau}}, \bibinfo {author}
  {\bibfnamefont {G.}~\bibnamefont {Fudenberg}}, \bibinfo {author}
  {\bibfnamefont {M.}~\bibnamefont {Imakaev}}, \bibinfo {author} {\bibfnamefont
  {L.~A.}\ \bibnamefont {Mirny}}, \bibinfo {author} {\bibfnamefont {C.-t.}\
  \bibnamefont {Wu}}, \ and\ \bibinfo {author} {\bibfnamefont {X.}~\bibnamefont
  {Zhuang}},\ }\href {\doibase 10.1038/nature16496} {\bibfield  {journal}
  {\bibinfo  {journal} {Nature}\ }\textbf {\bibinfo {volume} {529}},\ \bibinfo
  {pages} {418} (\bibinfo {year} {2016})}\BibitemShut {NoStop}%
\bibitem [{\citenamefont {Brackley}\ \emph {et~al.}(2013)\citenamefont
  {Brackley}, \citenamefont {Taylor}, \citenamefont {Papantonis}, \citenamefont
  {Cook},\ and\ \citenamefont {Marenduzzo}}]{Brackley2013pnas}%
  \BibitemOpen
  \bibfield  {author} {\bibinfo {author} {\bibfnamefont {C.~A.}\ \bibnamefont
  {Brackley}}, \bibinfo {author} {\bibfnamefont {S.}~\bibnamefont {Taylor}},
  \bibinfo {author} {\bibfnamefont {A.}~\bibnamefont {Papantonis}}, \bibinfo
  {author} {\bibfnamefont {P.~R.}\ \bibnamefont {Cook}}, \ and\ \bibinfo
  {author} {\bibfnamefont {D.}~\bibnamefont {Marenduzzo}},\ }\href {\doibase
  10.1073/pnas.1302950110} {\bibfield  {journal} {\bibinfo  {journal} {Proc.
  Natl. Acad. Sci. USA}\ }\textbf {\bibinfo {volume} {110}},\ \bibinfo {pages}
  {E3605} (\bibinfo {year} {2013})}\BibitemShut {NoStop}%
\bibitem [{\citenamefont {Brackley}\ \emph
  {et~al.}(2016{\natexlab{a}})\citenamefont {Brackley}, \citenamefont
  {Johnson}, \citenamefont {Kelly}, \citenamefont {Cook},\ and\ \citenamefont
  {Marenduzzo}}]{Brackley2016nar}%
  \BibitemOpen
  \bibfield  {author} {\bibinfo {author} {\bibfnamefont {C.~A.}\ \bibnamefont
  {Brackley}}, \bibinfo {author} {\bibfnamefont {J.}~\bibnamefont {Johnson}},
  \bibinfo {author} {\bibfnamefont {S.}~\bibnamefont {Kelly}}, \bibinfo
  {author} {\bibfnamefont {P.~R.}\ \bibnamefont {Cook}}, \ and\ \bibinfo
  {author} {\bibfnamefont {D.}~\bibnamefont {Marenduzzo}},\ }\href {\doibase
  10.1093/nar/gkw135} {\bibfield  {journal} {\bibinfo  {journal} {Nucleic Acids
  Res.}\ }\textbf {\bibinfo {volume} {44}},\ \bibinfo {pages} {3503} (\bibinfo
  {year} {2016}{\natexlab{a}})}\BibitemShut {NoStop}%
\bibitem [{\citenamefont {Brackley}\ \emph
  {et~al.}(2016{\natexlab{b}})\citenamefont {Brackley}, \citenamefont {Brown},
  \citenamefont {Waithe}, \citenamefont {Babbs}, \citenamefont {Davies},
  \citenamefont {Hughes}, \citenamefont {Buckle},\ and\ \citenamefont
  {Marenduzzo}}]{Brackley2016genomebiol}%
  \BibitemOpen
  \bibfield  {author} {\bibinfo {author} {\bibfnamefont {C.~A.}\ \bibnamefont
  {Brackley}}, \bibinfo {author} {\bibfnamefont {J.~M.}\ \bibnamefont {Brown}},
  \bibinfo {author} {\bibfnamefont {D.}~\bibnamefont {Waithe}}, \bibinfo
  {author} {\bibfnamefont {C.}~\bibnamefont {Babbs}}, \bibinfo {author}
  {\bibfnamefont {J.}~\bibnamefont {Davies}}, \bibinfo {author} {\bibfnamefont
  {J.~R.}\ \bibnamefont {Hughes}}, \bibinfo {author} {\bibfnamefont {V.~J.}\
  \bibnamefont {Buckle}}, \ and\ \bibinfo {author} {\bibfnamefont
  {D.}~\bibnamefont {Marenduzzo}},\ }\href {\doibase 10.1186/s13059-016-0909-0}
  {\bibfield  {journal} {\bibinfo  {journal} {Genome Biol.}\ }\textbf {\bibinfo
  {volume} {17}},\ \bibinfo {pages} {31} (\bibinfo {year}
  {2016}{\natexlab{b}})}\BibitemShut {NoStop}%
\bibitem [{\citenamefont {Barbieri}\ \emph {et~al.}(2012)\citenamefont
  {Barbieri}, \citenamefont {Chotalia}, \citenamefont {Fraser}, \citenamefont
  {Lavitas}, \citenamefont {Dostie}, \citenamefont {Pombo},\ and\ \citenamefont
  {Nicodemi}}]{Barbieri2012}%
  \BibitemOpen
  \bibfield  {author} {\bibinfo {author} {\bibfnamefont {M.}~\bibnamefont
  {Barbieri}}, \bibinfo {author} {\bibfnamefont {M.}~\bibnamefont {Chotalia}},
  \bibinfo {author} {\bibfnamefont {J.}~\bibnamefont {Fraser}}, \bibinfo
  {author} {\bibfnamefont {L.-M.}\ \bibnamefont {Lavitas}}, \bibinfo {author}
  {\bibfnamefont {J.}~\bibnamefont {Dostie}}, \bibinfo {author} {\bibfnamefont
  {A.}~\bibnamefont {Pombo}}, \ and\ \bibinfo {author} {\bibfnamefont
  {M.}~\bibnamefont {Nicodemi}},\ }\href {\doibase 10.1073/pnas.1204799109}
  {\bibfield  {journal} {\bibinfo  {journal} {Proc. Natl. Acad. Sci. USA}\
  }\textbf {\bibinfo {volume} {109}},\ \bibinfo {pages} {16173} (\bibinfo
  {year} {2012})}\BibitemShut {NoStop}%
\bibitem [{\citenamefont {Jost}\ \emph {et~al.}(2014)\citenamefont {Jost},
  \citenamefont {Carrivain}, \citenamefont {Cavalli},\ and\ \citenamefont
  {Vaillant}}]{Jost2014B}%
  \BibitemOpen
  \bibfield  {author} {\bibinfo {author} {\bibfnamefont {D.}~\bibnamefont
  {Jost}}, \bibinfo {author} {\bibfnamefont {P.}~\bibnamefont {Carrivain}},
  \bibinfo {author} {\bibfnamefont {G.}~\bibnamefont {Cavalli}}, \ and\
  \bibinfo {author} {\bibfnamefont {C.}~\bibnamefont {Vaillant}},\ }\href
  {\doibase 10.1093/nar/gku698} {\bibfield  {journal} {\bibinfo  {journal}
  {Nucleic Acids Res.}\ }\textbf {\bibinfo {volume} {42}},\ \bibinfo {pages}
  {1} (\bibinfo {year} {2014})}\BibitemShut {NoStop}%
\bibitem [{\citenamefont {Dodd}\ \emph {et~al.}(2007)\citenamefont {Dodd},
  \citenamefont {Micheelsen}, \citenamefont {Sneppen},\ and\ \citenamefont
  {Thon}}]{Dodd2007}%
  \BibitemOpen
  \bibfield  {author} {\bibinfo {author} {\bibfnamefont {I.~B.}\ \bibnamefont
  {Dodd}}, \bibinfo {author} {\bibfnamefont {M.~a.}\ \bibnamefont
  {Micheelsen}}, \bibinfo {author} {\bibfnamefont {K.}~\bibnamefont {Sneppen}},
  \ and\ \bibinfo {author} {\bibfnamefont {G.}~\bibnamefont {Thon}},\ }\href
  {\doibase 10.1016/j.cell.2007.02.053} {\bibfield  {journal} {\bibinfo
  {journal} {Cell}\ }\textbf {\bibinfo {volume} {129}},\ \bibinfo {pages} {813}
  (\bibinfo {year} {2007})}\BibitemShut {NoStop}%
\bibitem [{\citenamefont {Micheelsen}\ \emph {et~al.}(2010)\citenamefont
  {Micheelsen}, \citenamefont {Mitarai}, \citenamefont {Sneppen},\ and\
  \citenamefont {Dodd}}]{Micheelsen2010}%
  \BibitemOpen
  \bibfield  {author} {\bibinfo {author} {\bibfnamefont {M.~A.}\ \bibnamefont
  {Micheelsen}}, \bibinfo {author} {\bibfnamefont {N.}~\bibnamefont {Mitarai}},
  \bibinfo {author} {\bibfnamefont {K.}~\bibnamefont {Sneppen}}, \ and\
  \bibinfo {author} {\bibfnamefont {I.~B.}\ \bibnamefont {Dodd}},\ }\href
  {\doibase 10.1088/1478-3975/7/2/026010} {\bibfield  {journal} {\bibinfo
  {journal} {Phys. Biol.}\ }\textbf {\bibinfo {volume} {7}},\ \bibinfo {pages}
  {026010} (\bibinfo {year} {2010})}\BibitemShut {NoStop}%
\bibitem [{\citenamefont {Jost}(2014)}]{Jost2014pre}%
  \BibitemOpen
  \bibfield  {author} {\bibinfo {author} {\bibfnamefont {D.}~\bibnamefont
  {Jost}},\ }\href {\doibase 10.1103/PhysRevE.89.010701} {\bibfield  {journal}
  {\bibinfo  {journal} {Phys. Rev. E}\ }\textbf {\bibinfo {volume} {89}},\
  \bibinfo {pages} {1} (\bibinfo {year} {2014})}\BibitemShut {NoStop}%
\bibitem [{\citenamefont {Berry}\ \emph {et~al.}(2017)\citenamefont {Berry},
  \citenamefont {Dean},\ and\ \citenamefont {Howard}}]{Berry2017}%
  \BibitemOpen
  \bibfield  {author} {\bibinfo {author} {\bibfnamefont {S.}~\bibnamefont
  {Berry}}, \bibinfo {author} {\bibfnamefont {C.}~\bibnamefont {Dean}}, \ and\
  \bibinfo {author} {\bibfnamefont {M.}~\bibnamefont {Howard}},\ }\href
  {\doibase 10.1016/j.cels.2017.02.013} {\bibfield  {journal} {\bibinfo
  {journal} {Cell Syst.}\ }\textbf {\bibinfo {volume} {4}},\ \bibinfo {pages}
  {445} (\bibinfo {year} {2017})}\BibitemShut {NoStop}%
\bibitem [{\citenamefont {Garel}\ \emph {et~al.}(1999)\citenamefont {Garel},
  \citenamefont {Orland},\ and\ \citenamefont {Orlandini}}]{Garel1999a}%
  \BibitemOpen
  \bibfield  {author} {\bibinfo {author} {\bibfnamefont {T.}~\bibnamefont
  {Garel}}, \bibinfo {author} {\bibfnamefont {H.}~\bibnamefont {Orland}}, \
  and\ \bibinfo {author} {\bibfnamefont {E.}~\bibnamefont {Orlandini}},\
  }\href@noop {} {\bibfield  {journal} {\bibinfo  {journal} {EPJ B}\ }\textbf
  {\bibinfo {volume} {268}},\ \bibinfo {pages} {261} (\bibinfo {year}
  {1999})}\BibitemShut {NoStop}%
\bibitem [{\citenamefont {Michieletto}\ \emph {et~al.}(2016)\citenamefont
  {Michieletto}, \citenamefont {Orlandini},\ and\ \citenamefont
  {Marenduzzo}}]{Michieletto2016prx}%
  \BibitemOpen
  \bibfield  {author} {\bibinfo {author} {\bibfnamefont {D.}~\bibnamefont
  {Michieletto}}, \bibinfo {author} {\bibfnamefont {E.}~\bibnamefont
  {Orlandini}}, \ and\ \bibinfo {author} {\bibfnamefont {D.}~\bibnamefont
  {Marenduzzo}},\ }\href {\doibase 10.1103/PhysRevX.6.041047} {\bibfield
  {journal} {\bibinfo  {journal} {Phys. Rev. X}\ }\textbf {\bibinfo {volume}
  {6}},\ \bibinfo {pages} {041047} (\bibinfo {year} {2016})}\BibitemShut
  {NoStop}%
\bibitem [{\citenamefont {Michieletto}\ \emph {et~al.}(2017)\citenamefont
  {Michieletto}, \citenamefont {Orlandini},\ and\ \citenamefont
  {Marenduzzo}}]{Michieletto2017scirep}%
  \BibitemOpen
  \bibfield  {author} {\bibinfo {author} {\bibfnamefont {D.}~\bibnamefont
  {Michieletto}}, \bibinfo {author} {\bibfnamefont {E.}~\bibnamefont
  {Orlandini}}, \ and\ \bibinfo {author} {\bibfnamefont {D.}~\bibnamefont
  {Marenduzzo}},\ }\href {\doibase 10.1038/s41598-017-13916-w} {\bibfield
  {journal} {\bibinfo  {journal} {Sci. Rep.}\ }\textbf {\bibinfo {volume}
  {7}},\ \bibinfo {pages} {14642} (\bibinfo {year} {2017})}\BibitemShut
  {NoStop}%
\bibitem [{\citenamefont {Michieletto}\ \emph {et~al.}(2018)\citenamefont
  {Michieletto}, \citenamefont {Chiang}, \citenamefont {Coli}, \citenamefont
  {Papantonis}, \citenamefont {Orlandini}, \citenamefont {Cook},\ and\
  \citenamefont {Marenduzzo}}]{Michieletto2018nar}%
  \BibitemOpen
  \bibfield  {author} {\bibinfo {author} {\bibfnamefont {D.}~\bibnamefont
  {Michieletto}}, \bibinfo {author} {\bibfnamefont {M.}~\bibnamefont {Chiang}},
  \bibinfo {author} {\bibfnamefont {D.}~\bibnamefont {Coli}}, \bibinfo {author}
  {\bibfnamefont {A.}~\bibnamefont {Papantonis}}, \bibinfo {author}
  {\bibfnamefont {E.}~\bibnamefont {Orlandini}}, \bibinfo {author}
  {\bibfnamefont {P.~R.}\ \bibnamefont {Cook}}, \ and\ \bibinfo {author}
  {\bibfnamefont {D.}~\bibnamefont {Marenduzzo}},\ }\href {\doibase
  10.1093/nar/gkx1200} {\bibfield  {journal} {\bibinfo  {journal} {Nucleic
  Acids Res.}\ }\textbf {\bibinfo {volume} {46}},\ \bibinfo {pages} {83}
  (\bibinfo {year} {2018})}\BibitemShut {NoStop}%
\bibitem [{\citenamefont {Haddad}\ \emph {et~al.}(2017)\citenamefont {Haddad},
  \citenamefont {Jost},\ and\ \citenamefont {Vaillant}}]{Haddad2017}%
  \BibitemOpen
  \bibfield  {author} {\bibinfo {author} {\bibfnamefont {N.}~\bibnamefont
  {Haddad}}, \bibinfo {author} {\bibfnamefont {D.}~\bibnamefont {Jost}}, \ and\
  \bibinfo {author} {\bibfnamefont {C.}~\bibnamefont {Vaillant}},\ }\href
  {\doibase 10.1007/s10577-016-9548-2} {\bibfield  {journal} {\bibinfo
  {journal} {Chromosome Research}\ }\textbf {\bibinfo {volume} {25}},\ \bibinfo
  {pages} {35} (\bibinfo {year} {2017})}\BibitemShut {NoStop}%
\bibitem [{\citenamefont {Jost}\ and\ \citenamefont
  {Vaillant}(2018)}]{Jost2018}%
  \BibitemOpen
  \bibfield  {author} {\bibinfo {author} {\bibfnamefont {D.}~\bibnamefont
  {Jost}}\ and\ \bibinfo {author} {\bibfnamefont {C.}~\bibnamefont
  {Vaillant}},\ }\href {\doibase 10.1093/nar/gky009} {\bibfield  {journal}
  {\bibinfo  {journal} {Nucleic Acids Res.}\ }\textbf {\bibinfo {volume}
  {46}},\ \bibinfo {pages} {2252} (\bibinfo {year} {2018})}\BibitemShut
  {NoStop}%
\bibitem [{\citenamefont {Michieletto}\ \emph {et~al.}(2019)\citenamefont
  {Michieletto}, \citenamefont {Coli}, \citenamefont {Marenduzzo},\ and\
  \citenamefont {Orlandini}}]{PRL}%
  \BibitemOpen
  \bibfield  {author} {\bibinfo {author} {\bibfnamefont {D.}~\bibnamefont
  {Michieletto}}, \bibinfo {author} {\bibfnamefont {D.}~\bibnamefont {Coli}},
  \bibinfo {author} {\bibfnamefont {D.}~\bibnamefont {Marenduzzo}}, \ and\
  \bibinfo {author} {\bibfnamefont {E.}~\bibnamefont {Orlandini}},\ }\href@noop
  {} {\bibfield  {journal} {\bibinfo  {journal} {Phys. Rev. Lett.}\ }\textbf
  {\bibinfo {volume} {xx}},\ \bibinfo {pages} {xx} (\bibinfo {year}
  {2019})}\BibitemShut {NoStop}%
\bibitem [{\citenamefont {Lieberman-Aiden}\ \emph {et~al.}(2009)\citenamefont
  {Lieberman-Aiden}, \citenamefont {van Berkum}, \citenamefont {Williams},
  \citenamefont {Imakaev}, \citenamefont {Ragoczy}, \citenamefont {Telling},
  \citenamefont {Amit}, \citenamefont {Lajoie}, \citenamefont {Sabo},
  \citenamefont {Dorschner}, \citenamefont {Sandstrom}, \citenamefont
  {Bernstein}, \citenamefont {Bender}, \citenamefont {Groudine}, \citenamefont
  {Gnirke}, \citenamefont {Stamatoyannopoulos}, \citenamefont {Mirny},
  \citenamefont {Lander},\ and\ \citenamefont {Dekker}}]{Lieberman-Aiden2009}%
  \BibitemOpen
  \bibfield  {author} {\bibinfo {author} {\bibfnamefont {E.}~\bibnamefont
  {Lieberman-Aiden}}, \bibinfo {author} {\bibfnamefont {N.~L.}\ \bibnamefont
  {van Berkum}}, \bibinfo {author} {\bibfnamefont {L.}~\bibnamefont
  {Williams}}, \bibinfo {author} {\bibfnamefont {M.}~\bibnamefont {Imakaev}},
  \bibinfo {author} {\bibfnamefont {T.}~\bibnamefont {Ragoczy}}, \bibinfo
  {author} {\bibfnamefont {A.}~\bibnamefont {Telling}}, \bibinfo {author}
  {\bibfnamefont {I.}~\bibnamefont {Amit}}, \bibinfo {author} {\bibfnamefont
  {B.~R.}\ \bibnamefont {Lajoie}}, \bibinfo {author} {\bibfnamefont {P.~J.}\
  \bibnamefont {Sabo}}, \bibinfo {author} {\bibfnamefont {M.~O.}\ \bibnamefont
  {Dorschner}}, \bibinfo {author} {\bibfnamefont {R.}~\bibnamefont
  {Sandstrom}}, \bibinfo {author} {\bibfnamefont {B.}~\bibnamefont
  {Bernstein}}, \bibinfo {author} {\bibfnamefont {M.~a.}\ \bibnamefont
  {Bender}}, \bibinfo {author} {\bibfnamefont {M.}~\bibnamefont {Groudine}},
  \bibinfo {author} {\bibfnamefont {A.}~\bibnamefont {Gnirke}}, \bibinfo
  {author} {\bibfnamefont {J.}~\bibnamefont {Stamatoyannopoulos}}, \bibinfo
  {author} {\bibfnamefont {L.~a.}\ \bibnamefont {Mirny}}, \bibinfo {author}
  {\bibfnamefont {E.~S.}\ \bibnamefont {Lander}}, \ and\ \bibinfo {author}
  {\bibfnamefont {J.}~\bibnamefont {Dekker}},\ }\href {\doibase
  10.1126/science.1181369} {\bibfield  {journal} {\bibinfo  {journal} {Science
  (80-. ).}\ }\textbf {\bibinfo {volume} {326}},\ \bibinfo {pages} {289}
  (\bibinfo {year} {2009})}\BibitemShut {NoStop}%
\bibitem [{\citenamefont {Filion}\ \emph {et~al.}(2010)\citenamefont {Filion},
  \citenamefont {van Bemmel}, \citenamefont {Braunschweig}, \citenamefont
  {Talhout}, \citenamefont {Kind}, \citenamefont {Ward}, \citenamefont
  {Brugman}, \citenamefont {de~Castro}, \citenamefont {Kerkhoven},
  \citenamefont {Bussemaker},\ and\ \citenamefont {van Steensel}}]{Filion2010}%
  \BibitemOpen
  \bibfield  {author} {\bibinfo {author} {\bibfnamefont {G.~J.}\ \bibnamefont
  {Filion}}, \bibinfo {author} {\bibfnamefont {J.~G.}\ \bibnamefont {van
  Bemmel}}, \bibinfo {author} {\bibfnamefont {U.}~\bibnamefont {Braunschweig}},
  \bibinfo {author} {\bibfnamefont {W.}~\bibnamefont {Talhout}}, \bibinfo
  {author} {\bibfnamefont {J.}~\bibnamefont {Kind}}, \bibinfo {author}
  {\bibfnamefont {L.~D.}\ \bibnamefont {Ward}}, \bibinfo {author}
  {\bibfnamefont {W.}~\bibnamefont {Brugman}}, \bibinfo {author} {\bibfnamefont
  {I.~J.}\ \bibnamefont {de~Castro}}, \bibinfo {author} {\bibfnamefont {R.~M.}\
  \bibnamefont {Kerkhoven}}, \bibinfo {author} {\bibfnamefont {H.~J.}\
  \bibnamefont {Bussemaker}}, \ and\ \bibinfo {author} {\bibfnamefont
  {B.}~\bibnamefont {van Steensel}},\ }\href {\doibase
  10.1016/j.cell.2010.09.009} {\bibfield  {journal} {\bibinfo  {journal}
  {Cell}\ }\textbf {\bibinfo {volume} {143}},\ \bibinfo {pages} {212} (\bibinfo
  {year} {2010})}\BibitemShut {NoStop}%
\bibitem [{\citenamefont {Larson}\ \emph {et~al.}(2017)\citenamefont {Larson},
  \citenamefont {Elnatan}, \citenamefont {Keenen}, \citenamefont {Trnka},
  \citenamefont {Johnston}, \citenamefont {Burlingame}, \citenamefont {Agard},
  \citenamefont {Redding},\ and\ \citenamefont {Narlikar}}]{Larson2017}%
  \BibitemOpen
  \bibfield  {author} {\bibinfo {author} {\bibfnamefont {A.~G.}\ \bibnamefont
  {Larson}}, \bibinfo {author} {\bibfnamefont {D.}~\bibnamefont {Elnatan}},
  \bibinfo {author} {\bibfnamefont {M.~M.}\ \bibnamefont {Keenen}}, \bibinfo
  {author} {\bibfnamefont {M.~J.}\ \bibnamefont {Trnka}}, \bibinfo {author}
  {\bibfnamefont {J.~B.}\ \bibnamefont {Johnston}}, \bibinfo {author}
  {\bibfnamefont {A.~L.}\ \bibnamefont {Burlingame}}, \bibinfo {author}
  {\bibfnamefont {D.~A.}\ \bibnamefont {Agard}}, \bibinfo {author}
  {\bibfnamefont {S.}~\bibnamefont {Redding}}, \ and\ \bibinfo {author}
  {\bibfnamefont {G.~J.}\ \bibnamefont {Narlikar}},\ }\href {\doibase
  10.1038/nature22822} {\bibfield  {journal} {\bibinfo  {journal} {Nature}\
  }\textbf {\bibinfo {volume} {547}},\ \bibinfo {pages} {236} (\bibinfo {year}
  {2017})}\BibitemShut {NoStop}%
\bibitem [{\citenamefont {Pinter}\ \emph {et~al.}(2012)\citenamefont {Pinter},
  \citenamefont {Sadreyev}, \citenamefont {Yildirim}, \citenamefont {Jeon},
  \citenamefont {Ohsumi}, \citenamefont {Borowsky},\ and\ \citenamefont
  {Lee}}]{Pinter2012}%
  \BibitemOpen
  \bibfield  {author} {\bibinfo {author} {\bibfnamefont {S.~F.}\ \bibnamefont
  {Pinter}}, \bibinfo {author} {\bibfnamefont {R.~I.}\ \bibnamefont
  {Sadreyev}}, \bibinfo {author} {\bibfnamefont {E.}~\bibnamefont {Yildirim}},
  \bibinfo {author} {\bibfnamefont {Y.}~\bibnamefont {Jeon}}, \bibinfo {author}
  {\bibfnamefont {T.~K.}\ \bibnamefont {Ohsumi}}, \bibinfo {author}
  {\bibfnamefont {M.}~\bibnamefont {Borowsky}}, \ and\ \bibinfo {author}
  {\bibfnamefont {J.~T.}\ \bibnamefont {Lee}},\ }\href {\doibase
  10.1101/gr.133751.111.1864} {\bibfield  {journal} {\bibinfo  {journal}
  {Genome Res.}\ }\textbf {\bibinfo {volume} {22}},\ \bibinfo {pages} {1864}
  (\bibinfo {year} {2012})}\BibitemShut {NoStop}%
\bibitem [{\citenamefont {Hathaway}\ \emph {et~al.}(2012)\citenamefont
  {Hathaway}, \citenamefont {Bell}, \citenamefont {Hodges}, \citenamefont
  {Miller}, \citenamefont {Neel},\ and\ \citenamefont
  {Crabtree}}]{Hathaway2012}%
  \BibitemOpen
  \bibfield  {author} {\bibinfo {author} {\bibfnamefont {N.~A.}\ \bibnamefont
  {Hathaway}}, \bibinfo {author} {\bibfnamefont {O.}~\bibnamefont {Bell}},
  \bibinfo {author} {\bibfnamefont {C.}~\bibnamefont {Hodges}}, \bibinfo
  {author} {\bibfnamefont {E.~L.}\ \bibnamefont {Miller}}, \bibinfo {author}
  {\bibfnamefont {D.~S.}\ \bibnamefont {Neel}}, \ and\ \bibinfo {author}
  {\bibfnamefont {G.~R.}\ \bibnamefont {Crabtree}},\ }\href {\doibase
  10.1016/j.cell.2012.03.052} {\bibfield  {journal} {\bibinfo  {journal}
  {Cell}\ }\textbf {\bibinfo {volume} {149}},\ \bibinfo {pages} {1447}
  (\bibinfo {year} {2012})}\BibitemShut {NoStop}%
\bibitem [{\citenamefont {Wu}(1982)}]{Wu1982}%
  \BibitemOpen
  \bibfield  {author} {\bibinfo {author} {\bibfnamefont {F.~Y.}\ \bibnamefont
  {Wu}},\ }\href {\doibase 10.1103/revmodphys.54.235} {\bibfield  {journal}
  {\bibinfo  {journal} {Reviews of Modern Physics}\ }\textbf {\bibinfo {volume}
  {54}},\ \bibinfo {pages} {235} (\bibinfo {year} {1982})}\BibitemShut
  {NoStop}%
\bibitem [{\citenamefont {Duplantier}(1987)}]{Duplantier1987}%
  \BibitemOpen
  \bibfield  {author} {\bibinfo {author} {\bibfnamefont {B.}~\bibnamefont
  {Duplantier}},\ }\href {http://link.aps.org/@@doi /10.1103/PhysRevB.35.5290}
  {\bibfield  {journal} {\bibinfo  {journal} {Phys. Rev. B}\ }\textbf {\bibinfo
  {volume} {35}},\ \bibinfo {pages} {5290} (\bibinfo {year}
  {1987})}\BibitemShut {NoStop}%
\bibitem [{\citenamefont {Orland}\ \emph {et~al.}(1985)\citenamefont {Orland},
  \citenamefont {Itzykson},\ and\ \citenamefont {de~Dominicis}}]{Orland1985}%
  \BibitemOpen
  \bibfield  {author} {\bibinfo {author} {\bibfnamefont {H.}~\bibnamefont
  {Orland}}, \bibinfo {author} {\bibfnamefont {C.}~\bibnamefont {Itzykson}}, \
  and\ \bibinfo {author} {\bibfnamefont {C.}~\bibnamefont {de~Dominicis}},\
  }\href {\doibase 10.1051/jphyslet:01985004608035300} {\bibfield  {journal}
  {\bibinfo  {journal} {Journal de Physique Lettres}\ }\textbf {\bibinfo
  {volume} {46}},\ \bibinfo {pages} {353} (\bibinfo {year} {1985})}\BibitemShut
  {NoStop}%
\bibitem [{\citenamefont {Nemirovsky}\ \emph {et~al.}(1992)\citenamefont
  {Nemirovsky}, \citenamefont {Dudowicz},\ and\ \citenamefont
  {Freed}}]{Nemirovsky1992}%
  \BibitemOpen
  \bibfield  {author} {\bibinfo {author} {\bibfnamefont {A.~M.}\ \bibnamefont
  {Nemirovsky}}, \bibinfo {author} {\bibfnamefont {J.}~\bibnamefont
  {Dudowicz}}, \ and\ \bibinfo {author} {\bibfnamefont {K.~F.}\ \bibnamefont
  {Freed}},\ }\href {\doibase 10.1007/bf01049041} {\bibfield  {journal}
  {\bibinfo  {journal} {Journal of Statistical Physics}\ }\textbf {\bibinfo
  {volume} {67}},\ \bibinfo {pages} {395} (\bibinfo {year} {1992})}\BibitemShut
  {NoStop}%
\bibitem [{\citenamefont {Nicodemi}\ and\ \citenamefont
  {Prisco}(2007)}]{Nicodemi2007}%
  \BibitemOpen
  \bibfield  {author} {\bibinfo {author} {\bibfnamefont {M.}~\bibnamefont
  {Nicodemi}}\ and\ \bibinfo {author} {\bibfnamefont {A.}~\bibnamefont
  {Prisco}},\ }\href {\doibase 10.1103/PhysRevLett.98.108104} {\bibfield
  {journal} {\bibinfo  {journal} {Phys. Rev. Lett.}\ }\textbf {\bibinfo
  {volume} {98}},\ \bibinfo {pages} {108104} (\bibinfo {year}
  {2007})}\BibitemShut {NoStop}%
\bibitem [{\citenamefont {Henikoff}\ and\ \citenamefont
  {Greally}(2016)}]{Henikoff2016}%
  \BibitemOpen
  \bibfield  {author} {\bibinfo {author} {\bibfnamefont {S.}~\bibnamefont
  {Henikoff}}\ and\ \bibinfo {author} {\bibfnamefont {J.~M.}\ \bibnamefont
  {Greally}},\ }\href {\doibase 10.1016/j.cub.2016.06.011} {\bibfield
  {journal} {\bibinfo  {journal} {Curr. Biol.}\ }\textbf {\bibinfo {volume}
  {26}},\ \bibinfo {pages} {R644} (\bibinfo {year} {2016})}\BibitemShut
  {NoStop}%
\bibitem [{\citenamefont {Michieletto}\ and\ \citenamefont
  {Gilbert}(2019)}]{Michieletto2019rnareview}%
  \BibitemOpen
  \bibfield  {author} {\bibinfo {author} {\bibfnamefont {D.}~\bibnamefont
  {Michieletto}}\ and\ \bibinfo {author} {\bibfnamefont {N.}~\bibnamefont
  {Gilbert}},\ }\href {\doibase 10.1016/j.ceb.2019.03.007} {\bibfield
  {journal} {\bibinfo  {journal} {Curr. Opin. Cell Biol.}\ }\textbf {\bibinfo
  {volume} {58}},\ \bibinfo {pages} {120} (\bibinfo {year} {2019})}\BibitemShut
  {NoStop}%
\bibitem [{\citenamefont {Festuccia}\ \emph {et~al.}(2016)\citenamefont
  {Festuccia}, \citenamefont {Dubois}, \citenamefont {Vandormael-Pournin},
  \citenamefont {Tejeda}, \citenamefont {Mouren}, \citenamefont {Bessonnard},
  \citenamefont {Mueller}, \citenamefont {Proux}, \citenamefont
  {Cohen-Tannoudji},\ and\ \citenamefont {Navarro}}]{Festuccia2016}%
  \BibitemOpen
  \bibfield  {author} {\bibinfo {author} {\bibfnamefont {N.}~\bibnamefont
  {Festuccia}}, \bibinfo {author} {\bibfnamefont {A.}~\bibnamefont {Dubois}},
  \bibinfo {author} {\bibfnamefont {S.}~\bibnamefont {Vandormael-Pournin}},
  \bibinfo {author} {\bibfnamefont {E.~G.}\ \bibnamefont {Tejeda}}, \bibinfo
  {author} {\bibfnamefont {A.}~\bibnamefont {Mouren}}, \bibinfo {author}
  {\bibfnamefont {S.}~\bibnamefont {Bessonnard}}, \bibinfo {author}
  {\bibfnamefont {F.}~\bibnamefont {Mueller}}, \bibinfo {author} {\bibfnamefont
  {C.}~\bibnamefont {Proux}}, \bibinfo {author} {\bibfnamefont
  {M.}~\bibnamefont {Cohen-Tannoudji}}, \ and\ \bibinfo {author} {\bibfnamefont
  {P.}~\bibnamefont {Navarro}},\ }\href {\doibase 10.1038/ncb3418} {\bibfield
  {journal} {\bibinfo  {journal} {Nat. Cell Biol.}\ }\textbf {\bibinfo {volume}
  {18}},\ \bibinfo {pages} {1139} (\bibinfo {year} {2016})}\BibitemShut
  {NoStop}%
\bibitem [{\citenamefont {Wang}\ \emph {et~al.}(2013)\citenamefont {Wang},
  \citenamefont {Liu},\ and\ \citenamefont {Dong}}]{Wang2013}%
  \BibitemOpen
  \bibfield  {author} {\bibinfo {author} {\bibfnamefont {S.}~\bibnamefont
  {Wang}}, \bibinfo {author} {\bibfnamefont {M.}~\bibnamefont {Liu}}, \ and\
  \bibinfo {author} {\bibfnamefont {Y.}~\bibnamefont {Dong}},\ }\href {\doibase
  10.1088/0953-8984/25/18/184007} {\bibfield  {journal} {\bibinfo  {journal}
  {J. Phys. Condens. Matter}\ }\textbf {\bibinfo {volume} {25}},\ \bibinfo
  {pages} {184007} (\bibinfo {year} {2013})}\BibitemShut {NoStop}%
\bibitem [{\citenamefont {Egli}\ \emph {et~al.}(2008)\citenamefont {Egli},
  \citenamefont {Birkhoff},\ and\ \citenamefont {Eggan}}]{Egli2008}%
  \BibitemOpen
  \bibfield  {author} {\bibinfo {author} {\bibfnamefont {D.}~\bibnamefont
  {Egli}}, \bibinfo {author} {\bibfnamefont {G.}~\bibnamefont {Birkhoff}}, \
  and\ \bibinfo {author} {\bibfnamefont {K.}~\bibnamefont {Eggan}},\ }\href
  {\doibase 10.1038/nrm2439} {\bibfield  {journal} {\bibinfo  {journal} {Nat.
  Rev. Mol. Cell. Biol.}\ }\textbf {\bibinfo {volume} {9}},\ \bibinfo {pages}
  {505} (\bibinfo {year} {2008})}\BibitemShut {NoStop}%
\bibitem [{\citenamefont {Hohenberg}\ and\ \citenamefont
  {Halperin}(1977)}]{Hohenberg1977}%
  \BibitemOpen
  \bibfield  {author} {\bibinfo {author} {\bibfnamefont {P.}~\bibnamefont
  {Hohenberg}}\ and\ \bibinfo {author} {\bibfnamefont {B.}~\bibnamefont
  {Halperin}},\ }\href {\doibase 10.1103/RevModPhys.49.435} {\bibfield
  {journal} {\bibinfo  {journal} {Rev. Mod. Phys.}\ }\textbf {\bibinfo {volume}
  {49}},\ \bibinfo {pages} {435} (\bibinfo {year} {1977})}\BibitemShut
  {NoStop}%
\bibitem [{\citenamefont {Chaikin}\ and\ \citenamefont
  {Lubensky}(2007)}]{ChaikinLubensky}%
  \BibitemOpen
  \bibfield  {author} {\bibinfo {author} {\bibfnamefont {P.~M.}\ \bibnamefont
  {Chaikin}}\ and\ \bibinfo {author} {\bibfnamefont {T.~C.}\ \bibnamefont
  {Lubensky}},\ }\href@noop {} {\emph {\bibinfo {title} {{Principles of
  Condensed Matter Physics}}}},\ Vol.~\bibinfo {volume} {c}\ (\bibinfo
  {publisher} {Cambridge University Press},\ \bibinfo {year}
  {2007})\BibitemShut {NoStop}%
\bibitem [{\citenamefont {Sexton}\ \emph {et~al.}(2012)\citenamefont {Sexton},
  \citenamefont {Yaffe}, \citenamefont {Kenigsberg}, \citenamefont
  {Bantignies}, \citenamefont {Leblanc}, \citenamefont {Hoichman},
  \citenamefont {Parrinello}, \citenamefont {Tanay},\ and\ \citenamefont
  {Cavalli}}]{Sexton2012}%
  \BibitemOpen
  \bibfield  {author} {\bibinfo {author} {\bibfnamefont {T.}~\bibnamefont
  {Sexton}}, \bibinfo {author} {\bibfnamefont {E.}~\bibnamefont {Yaffe}},
  \bibinfo {author} {\bibfnamefont {E.}~\bibnamefont {Kenigsberg}}, \bibinfo
  {author} {\bibfnamefont {F.}~\bibnamefont {Bantignies}}, \bibinfo {author}
  {\bibfnamefont {B.}~\bibnamefont {Leblanc}}, \bibinfo {author} {\bibfnamefont
  {M.}~\bibnamefont {Hoichman}}, \bibinfo {author} {\bibfnamefont
  {H.}~\bibnamefont {Parrinello}}, \bibinfo {author} {\bibfnamefont
  {A.}~\bibnamefont {Tanay}}, \ and\ \bibinfo {author} {\bibfnamefont
  {G.}~\bibnamefont {Cavalli}},\ }\href {\doibase 10.1016/j.cell.2012.01.010}
  {\bibfield  {journal} {\bibinfo  {journal} {Cell}\ }\textbf {\bibinfo
  {volume} {148}},\ \bibinfo {pages} {458} (\bibinfo {year}
  {2012})}\BibitemShut {NoStop}%
\bibitem [{\citenamefont {Szabo}\ \emph {et~al.}(2018)\citenamefont {Szabo},
  \citenamefont {Jost}, \citenamefont {Chang}, \citenamefont {Cattoni},
  \citenamefont {Papadopoulos}, \citenamefont {Bonev}, \citenamefont {Sexton},
  \citenamefont {Gurgo}, \citenamefont {Jacquier}, \citenamefont {Nollmann},
  \citenamefont {Bantignies},\ and\ \citenamefont {Cavalli}}]{Szabo2018}%
  \BibitemOpen
  \bibfield  {author} {\bibinfo {author} {\bibfnamefont {Q.}~\bibnamefont
  {Szabo}}, \bibinfo {author} {\bibfnamefont {D.}~\bibnamefont {Jost}},
  \bibinfo {author} {\bibfnamefont {J.~M.}\ \bibnamefont {Chang}}, \bibinfo
  {author} {\bibfnamefont {D.~I.}\ \bibnamefont {Cattoni}}, \bibinfo {author}
  {\bibfnamefont {G.~L.}\ \bibnamefont {Papadopoulos}}, \bibinfo {author}
  {\bibfnamefont {B.}~\bibnamefont {Bonev}}, \bibinfo {author} {\bibfnamefont
  {T.}~\bibnamefont {Sexton}}, \bibinfo {author} {\bibfnamefont
  {J.}~\bibnamefont {Gurgo}}, \bibinfo {author} {\bibfnamefont
  {C.}~\bibnamefont {Jacquier}}, \bibinfo {author} {\bibfnamefont
  {M.}~\bibnamefont {Nollmann}}, \bibinfo {author} {\bibfnamefont
  {F.}~\bibnamefont {Bantignies}}, \ and\ \bibinfo {author} {\bibfnamefont
  {G.}~\bibnamefont {Cavalli}},\ }\href {\doibase 10.1126/sciadv.aar8082}
  {\bibfield  {journal} {\bibinfo  {journal} {Sci. Adv.}\ }\textbf {\bibinfo
  {volume} {4}},\ \bibinfo {pages} {1} (\bibinfo {year} {2018})}\BibitemShut
  {NoStop}%
\bibitem [{\citenamefont {Rosa}\ and\ \citenamefont
  {Everaers}(2008)}]{Rosa2008}%
  \BibitemOpen
  \bibfield  {author} {\bibinfo {author} {\bibfnamefont {A.}~\bibnamefont
  {Rosa}}\ and\ \bibinfo {author} {\bibfnamefont {R.}~\bibnamefont
  {Everaers}},\ }\href {\doibase 10.1371/journal.pcbi.1000153} {\bibfield
  {journal} {\bibinfo  {journal} {PLoS Comp. Biol.}\ }\textbf {\bibinfo
  {volume} {4}},\ \bibinfo {pages} {1} (\bibinfo {year} {2008})}\BibitemShut
  {NoStop}%
\bibitem [{\citenamefont {Kremer}\ and\ \citenamefont
  {Grest}(1990)}]{Kremer1990}%
  \BibitemOpen
  \bibfield  {author} {\bibinfo {author} {\bibfnamefont {K.}~\bibnamefont
  {Kremer}}\ and\ \bibinfo {author} {\bibfnamefont {G.~S.}\ \bibnamefont
  {Grest}},\ }\href {\doibase 10.1063/1.458541} {\bibfield  {journal} {\bibinfo
   {journal} {J. Chem. Phys.}\ }\textbf {\bibinfo {volume} {92}},\ \bibinfo
  {pages} {5057} (\bibinfo {year} {1990})}\BibitemShut {NoStop}%
\bibitem [{\citenamefont {Socol}\ \emph {et~al.}(2019)\citenamefont {Socol},
  \citenamefont {Wang}, \citenamefont {Jost}, \citenamefont {Carrivain},
  \citenamefont {Vaillant}, \citenamefont {{Le Cam}}, \citenamefont {Dahirel},
  \citenamefont {Normand}, \citenamefont {Bystricky}, \citenamefont {Victor},
  \citenamefont {Gadal},\ and\ \citenamefont {Bancaud}}]{Socol2019}%
  \BibitemOpen
  \bibfield  {author} {\bibinfo {author} {\bibfnamefont {M.}~\bibnamefont
  {Socol}}, \bibinfo {author} {\bibfnamefont {R.}~\bibnamefont {Wang}},
  \bibinfo {author} {\bibfnamefont {D.}~\bibnamefont {Jost}}, \bibinfo {author}
  {\bibfnamefont {P.}~\bibnamefont {Carrivain}}, \bibinfo {author}
  {\bibfnamefont {C.}~\bibnamefont {Vaillant}}, \bibinfo {author}
  {\bibfnamefont {E.}~\bibnamefont {{Le Cam}}}, \bibinfo {author}
  {\bibfnamefont {V.}~\bibnamefont {Dahirel}}, \bibinfo {author} {\bibfnamefont
  {C.}~\bibnamefont {Normand}}, \bibinfo {author} {\bibfnamefont
  {K.}~\bibnamefont {Bystricky}}, \bibinfo {author} {\bibfnamefont {J.-M.}\
  \bibnamefont {Victor}}, \bibinfo {author} {\bibfnamefont {O.}~\bibnamefont
  {Gadal}}, \ and\ \bibinfo {author} {\bibfnamefont {A.}~\bibnamefont
  {Bancaud}},\ }\href {\doibase 10.1093/nar/gkz374} {\bibfield  {journal}
  {\bibinfo  {journal} {Nucleic Acids Research}\ }\textbf {\bibinfo {volume}
  {47}},\ \bibinfo {pages} {6195} (\bibinfo {year} {2019})}\BibitemShut
  {NoStop}%
\bibitem [{\citenamefont {Plimpton}(1995)}]{Plimpton1995}%
  \BibitemOpen
  \bibfield  {author} {\bibinfo {author} {\bibfnamefont {S.}~\bibnamefont
  {Plimpton}},\ }\href {\doibase 10.1006/jcph.1995.1039} {\bibfield  {journal}
  {\bibinfo  {journal} {J. Comp. Phys.}\ }\textbf {\bibinfo {volume} {117}},\
  \bibinfo {pages} {1} (\bibinfo {year} {1995})}\BibitemShut {NoStop}%
\bibitem [{\citenamefont {Brackley}\ \emph
  {et~al.}(2017{\natexlab{a}})\citenamefont {Brackley}, \citenamefont
  {Liebchen}, \citenamefont {Michieletto}, \citenamefont {Mouvet},
  \citenamefont {Cook},\ and\ \citenamefont
  {Marenduzzo}}]{Brackley2017biophysj}%
  \BibitemOpen
  \bibfield  {author} {\bibinfo {author} {\bibfnamefont {C.~A.}\ \bibnamefont
  {Brackley}}, \bibinfo {author} {\bibfnamefont {B.}~\bibnamefont {Liebchen}},
  \bibinfo {author} {\bibfnamefont {D.}~\bibnamefont {Michieletto}}, \bibinfo
  {author} {\bibfnamefont {F.}~\bibnamefont {Mouvet}}, \bibinfo {author}
  {\bibfnamefont {P.~R.}\ \bibnamefont {Cook}}, \ and\ \bibinfo {author}
  {\bibfnamefont {D.}~\bibnamefont {Marenduzzo}},\ }\href {\doibase
  10.1016/j.bpj.2017.01.025} {\bibfield  {journal} {\bibinfo  {journal}
  {Biophys J.}\ }\textbf {\bibinfo {volume} {112}},\ \bibinfo {pages} {1085}
  (\bibinfo {year} {2017}{\natexlab{a}})}\BibitemShut {NoStop}%
\bibitem [{\citenamefont {Orlandini}\ \emph {et~al.}(2019)\citenamefont
  {Orlandini}, \citenamefont {Marenduzzo},\ and\ \citenamefont
  {Michieletto}}]{Orlandini2019}%
  \BibitemOpen
  \bibfield  {author} {\bibinfo {author} {\bibfnamefont {E.}~\bibnamefont
  {Orlandini}}, \bibinfo {author} {\bibfnamefont {D.}~\bibnamefont
  {Marenduzzo}}, \ and\ \bibinfo {author} {\bibfnamefont {D.}~\bibnamefont
  {Michieletto}},\ }\href {\doibase 10.1073/pnas.1815394116} {\bibfield
  {journal} {\bibinfo  {journal} {Proc. Natl. Acad. Sci.}\ ,\ \bibinfo {pages}
  {201815394}} (\bibinfo {year} {2019})}\BibitemShut {NoStop}%
\bibitem [{\citenamefont {Fudenberg}\ \emph {et~al.}(2016)\citenamefont
  {Fudenberg}, \citenamefont {Imakaev}, \citenamefont {Lu}, \citenamefont
  {Goloborodko}, \citenamefont {Abdennur},\ and\ \citenamefont
  {Mirny}}]{Fudenberg2016}%
  \BibitemOpen
  \bibfield  {author} {\bibinfo {author} {\bibfnamefont {G.}~\bibnamefont
  {Fudenberg}}, \bibinfo {author} {\bibfnamefont {M.}~\bibnamefont {Imakaev}},
  \bibinfo {author} {\bibfnamefont {C.}~\bibnamefont {Lu}}, \bibinfo {author}
  {\bibfnamefont {A.}~\bibnamefont {Goloborodko}}, \bibinfo {author}
  {\bibfnamefont {N.}~\bibnamefont {Abdennur}}, \ and\ \bibinfo {author}
  {\bibfnamefont {L.~A.}\ \bibnamefont {Mirny}},\ }\href {\doibase
  10.1101/024620} {\bibfield  {journal} {\bibinfo  {journal} {Cell Rep.}\
  }\textbf {\bibinfo {volume} {15}},\ \bibinfo {pages} {2038} (\bibinfo {year}
  {2016})}\BibitemShut {NoStop}%
\bibitem [{\citenamefont {Brackley}\ \emph
  {et~al.}(2017{\natexlab{b}})\citenamefont {Brackley}, \citenamefont
  {Johnson}, \citenamefont {Michieletto}, \citenamefont {Morozov},
  \citenamefont {Nicodemi}, \citenamefont {Cook},\ and\ \citenamefont
  {Marenduzzo}}]{Brackley2017prl}%
  \BibitemOpen
  \bibfield  {author} {\bibinfo {author} {\bibfnamefont {C.}~\bibnamefont
  {Brackley}}, \bibinfo {author} {\bibfnamefont {J.}~\bibnamefont {Johnson}},
  \bibinfo {author} {\bibfnamefont {D.}~\bibnamefont {Michieletto}}, \bibinfo
  {author} {\bibfnamefont {A.}~\bibnamefont {Morozov}}, \bibinfo {author}
  {\bibfnamefont {M.}~\bibnamefont {Nicodemi}}, \bibinfo {author}
  {\bibfnamefont {P.}~\bibnamefont {Cook}}, \ and\ \bibinfo {author}
  {\bibfnamefont {D.}~\bibnamefont {Marenduzzo}},\ }\href {\doibase
  10.1103/PhysRevLett.119.138101} {\bibfield  {journal} {\bibinfo  {journal}
  {Phys. Rev. Lett.}\ }\textbf {\bibinfo {volume} {119}},\ \bibinfo {pages}
  {138101} (\bibinfo {year} {2017}{\natexlab{b}})}\BibitemShut {NoStop}%
\bibitem [{\citenamefont {Tauran}\ \emph {et~al.}(2019)\citenamefont {Tauran},
  \citenamefont {Kumemura}, \citenamefont {Tarhan}, \citenamefont {Perret},
  \citenamefont {Perret}, \citenamefont {Jalabert}, \citenamefont {Collard},
  \citenamefont {Fujita},\ and\ \citenamefont {Coleman}}]{Tauran2019}%
  \BibitemOpen
  \bibfield  {author} {\bibinfo {author} {\bibfnamefont {Y.}~\bibnamefont
  {Tauran}}, \bibinfo {author} {\bibfnamefont {M.}~\bibnamefont {Kumemura}},
  \bibinfo {author} {\bibfnamefont {M.~C.}\ \bibnamefont {Tarhan}}, \bibinfo
  {author} {\bibfnamefont {G.}~\bibnamefont {Perret}}, \bibinfo {author}
  {\bibfnamefont {F.}~\bibnamefont {Perret}}, \bibinfo {author} {\bibfnamefont
  {L.}~\bibnamefont {Jalabert}}, \bibinfo {author} {\bibfnamefont
  {D.}~\bibnamefont {Collard}}, \bibinfo {author} {\bibfnamefont
  {H.}~\bibnamefont {Fujita}}, \ and\ \bibinfo {author} {\bibfnamefont {A.~W.}\
  \bibnamefont {Coleman}},\ }\href {\doibase 10.1038/s41598-019-42267-x}
  {\bibfield  {journal} {\bibinfo  {journal} {Scientific Reports}\ }\textbf
  {\bibinfo {volume} {9}},\ \bibinfo {pages} {1} (\bibinfo {year}
  {2019})}\BibitemShut {NoStop}%
\end{thebibliography}%

\end{document}